\documentclass[pdflatex,sn-basic,iicol]{sn-jnl}

\usepackage{graphicx}%
\usepackage{multirow}%
\usepackage{amsmath,amssymb,amsfonts}%
\usepackage{amsthm}%
\usepackage{mathrsfs}%
\usepackage[title]{appendix}%
\usepackage{xcolor}%
\usepackage{textcomp}%
\usepackage{manyfoot}%
\usepackage{booktabs}%
\usepackage{algorithm}%
\usepackage{algorithmicx}%
\usepackage{algpseudocode}%
\usepackage{listings}%

\unnumbered

\begin{document}

\title[Emergence of Functionally Differentiated Structures via Mutual Information Minimization in Recurrent Neural Networks]{Emergence of Functionally Differentiated Structures via Mutual Information Minimization in Recurrent Neural Networks}


\author[1]{\fnm{Yuki} \sur{Tomoda}}\email{mfm24110@bene.fit.ac.jp}

\author[2]{\fnm{Ichiro} \sur{Tsuda}}\email{i.tsuda@scu.ac.jp}

\author*[3]{\fnm{Yutaka} \sur{Yamaguti}}\email{y-yamaguchi@fit.ac.jp}

\affil[1]{\orgdiv{Graduate School of Engineering}, \orgname{Fukuoka Institute of Technology}, \orgaddress{\street{3-30-1 Wajiro-Higashi, Higashi-ku}, \city{Fukuoka}, \state{Fukuoka} \postcode{811-0295}, \country{Japan}}}

\affil[2]{\orgdiv{AIT Center}, \orgname{Sapporo City University}, \orgaddress{\street{20-1, Minami 1 Nishi 6}, \city{Sapporo}, \state{Hokkaido} \postcode{060-0061}, \country{Japan}}}

\affil*[3]{\orgdiv{Faculty of Information Engineering}, \orgname{Fukuoka Institute of Technology}, \orgaddress{\street{3-30-1 Wajiro-Higashi, Higashi-ku}, \city{Fukuoka}, \state{Fukuoka} \postcode{811-0295}, \country{Japan}}}

\abstract{Functional differentiation in the brain emerges as distinct regions specialize and is key to understanding brain function as a complex system. Previous research has modeled this process using artificial neural networks with specific constraints. Here, we propose a novel approach that induces functional differentiation in recurrent neural networks by minimizing mutual information between neural subgroups via mutual information neural estimation. We apply our method to a 2-bit working memory task and a chaotic signal separation task involving Lorenz and R\"ossler time series. Analysis of network performance, correlation patterns, and weight matrices reveals that mutual information minimization yields high task performance alongside clear functional modularity and moderate structural modularity. Importantly, our results show that functional differentiation, which is measured through correlation structures, emerges earlier than structural modularity defined by synaptic weights. This suggests that functional specialization precedes and probably drives structural reorganization within developing neural networks. Our findings provide new insights into how information-theoretic principles may govern the emergence of specialized functions and modular structures during artificial and biological brain development.}


\maketitle

\section{Introduction}\label{sec1}

During brain development, populations of neurons with specific functions organize into different regions, forming a functional map, in a process called functional differentiation~\citep{brodmann1909vergleichende}.
Studies analyzing finer structural division called parcellation have revealed that these functional areas are dynamically reorganized according to specific tasks~\citep{Glasser2016}.
Throughout these developmental processes, modular structures consistently emerge in the brain's organization~\citep{felleman1991,Glasser2016,sporns2016,sporns2016book}.

However, the underlying principles driving the formation of these differentiated structures remain largely unknown~\citep{clune2013,sporns2016}. 
Thus, it is particularly valuable to study these problems to understand the mechanisms of emerging brain functions.
Mathematical modeling offers a valuable approach to formulate these problems quantitatively and test hypotheses systematically~\citep{kaneko2001book}.

Functional differentiation in the brain is also believed to be fundamental to the emergence of intelligence~\citep{kashtan2005,sporns2016,yang2019task}. 
To investigate the underlying computational principles of this process, we need models that capture both the temporal dynamics of neural activity and the capacity for adaptive specialization. 
Therefore, insights from recent advances in machine learning can be particularly relevant~\citep{sussillo2014neural,yang2019task}.

Recurrent neural networks (RNNs) provide an ideal framework for investigating functional differentiation. They exhibit temporal dynamics similar to biological circuits, have a long history in computational neuroscience~\citep{caianiello1961,amari1972,hopfield1982,elman1990}, and recent developments enable modeling of complex neural dynamics~\citep{sussillo2014neural,song2016training,yang2019task,yamaguti2021functional}.
We use RNNs to examine how functional differentiation emerges under information-theoretic constraints.

Previous research has proposed that functional differentiation can be understood within the framework of self-organization with constraints, where system components emerge under constraints that act on the entire system~\citep{tsuda2016,tsuda2022nature}. This framework suggests that the overall development may follow certain variational principles, and thus the differentiation process is not genetically predetermined. Instead, the components emerge dynamically as the system optimizes itself under global constraints.
Considering that the brain's primary information processing function is to acquire information from the environment to produce appropriate behaviors for survival, a compelling candidate for such a universal constraint is the optimization of information flow within neural networks.
Mutual information (MI) provides a quantitative measure for analyzing this information flow because it captures the statistical dependencies among different neural populations. However, calculating MI in complex neural systems presents major challenges, particularly when assessing dependencies among multiple variables in high-dimensional spaces.

To implement this theoretical framework, we use mutual information neural estimation (MINE)~\citep{belghazi18mine}, which enables efficient estimation of MI and gradient-based optimization in high-dimensional spaces. We propose inducing functional differentiation by minimizing MI between predefined neural subgroups. While previous studies maximized information-theoretic measures~\citep{linsker1988,bell1995,tanaka2009,watanabe2020mathematical,yamaguti2021functional}, we hypothesize that constraining subgroups to be statistically independent drives them to specialize in different task aspects without explicitly specifying their functions.

\begin{figure*}[tbph]
    \centering 
     \includegraphics[width=\textwidth]{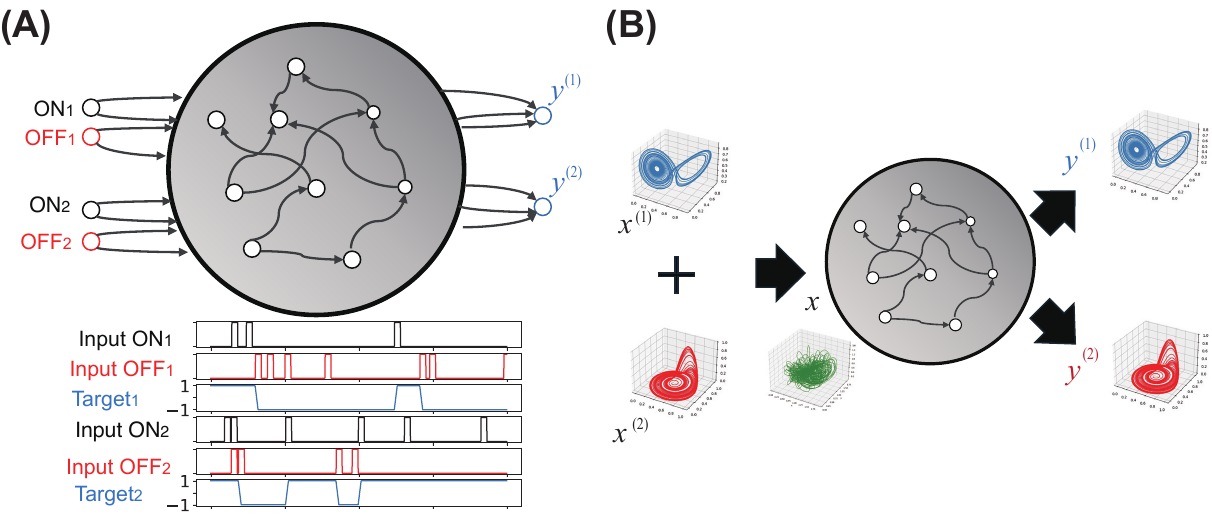} 
     \caption{Schematics of the two experimental tasks. (A) Working memory task. The network receives brief pulses on four input channels (ON$_1$, OFF$_1$, ON$_2$, OFF$_2$) and must maintain two independent memory bits (output values of $\pm 1.0$) based on the most recent pulse received. (B) Chaotic signal separation task. The network receives a mixed 3-dimensional input signal from Lorenz and R\"ossler chaotic systems and must separate it into two distinct 3-dimensional output signals}
    \label{fig:tasks}   
\end{figure*}

We test our approach on two complementary tasks (Fig.~\ref{fig:tasks}): (1) a 2-bit working memory task~\citep{sussillo2009,hoerzer_2014} requiring independent maintenance of two binary states, and (2) separation of mixed chaotic time series from Lorenz and R\"ossler systems~\citep{lorenz1963,rossler1976}. Both tasks require specialized neural functions analyzable through our information-theoretic framework.

The remainder of this paper presents related work, methods, results from both tasks, and discussion of implications.

\section{Background}\label{sec:related} 
\subsection{Driving forces for functional differentiation}
Understanding the process through which cell populations differentiate into distinct groups with specific functions, as well as the evolutionary driving forces behind this process, is a key area in the study of complex systems~\citep{kaneko2001book,kaneko2006life}.
The question of how differentiation proceeds and why it is crucial for biological systems has
been discussed extensively in neuroscience, evolutionary developmental biology, and artificial intelligence~\citep{kashtan2005,espinosa-soto2010,clune2013,ellefsen2015,sporns2016book}.
Functional differentiation provides major advantages to biological systems, including greater robustness and higher connection efficiency, where modular organization enables effective communication with fewer and shorter connections~\citep{latora2001,sporns2016book}, better integration of information across different timescales~\citep{ichikawa2024}, and facilitated information transfer~\citep{yamaguti2015mathematical,yamaguti2021functional}.
The crucial role of modular structure in neurodynamics has been investigated in several studies~\citep{lord2017understanding,kawai2023learning}. 

We focus on information transfer as a key constraint for differentiation. 
\cite{yamaguti2015mathematical} demonstrated that coupled oscillator systems 
evolved to maximize the transfer entropy between subnetworks, which was 
accompanied by the emergence of asymmetric connections within these subnetworks, 
thus facilitating their differentiation. 
The precise mechanism underlying this relationship between inter-subnetwork 
information transfer and asymmetric connectivity remains unclear and may involve 
spontaneous symmetry breaking during the optimization process. 
These constraints that act as driving forces of differentiation are complementary rather than mutually exclusive, potentially working in concert to shape biological systems.

The approaches for investigating the emergence of functional differentiation in artificial neural networks can be broadly classified as bottom-up or top-down.

Bottom-up approaches either impose constraints only on local structures, such as synapses between neurons~\citep{malsburg1973,amari1980,kohonen1982}, or rely on the emergence of differentiation naturally through the requirements of the task itself~\citep{yang2019task}. 
For example, \cite{yang2019task} demonstrated how functionally specialized neural clusters emerge through training a simple RNN on multiple cognitive tasks.
By analyzing individual units and the effect of lesions, they revealed how complex cognitive abilities can arise from the interactions of neurons without explicitly imposing global organizational constraints.

In contrast, top-down approaches consider the existence of universal and global constraints that align with the general goals of biological systems~\citep{tsuda2016}. In our study, we adopt this top-down perspective through a variational principle that guides functional differentiation by imposing system-wide constraints~\citep{tsuda2016,tsuda2022nature,watanabe2020mathematical}.

\subsection{Information-theoretic approaches to neural network optimization}

In neural network learning, various optimization methods based on information-theoretic measures have been proposed. \cite{linsker1988} pioneered this approach with the InfoMax principle, which was applied to signal separation tasks by maximizing the MI between input and output~\citep{bell1995}. Building on this foundation, \cite{tanaka2009} later developed the concept of Recurrent InfoMax, adapting these principles specifically for RNNs.  

Several studies have employed evolutionary algorithms to optimize information flow in neural networks. 
\cite{watanabe2020mathematical} evolved networks of dynamic elements by maximizing time-dependent MI between network elements and inputs, demonstrating that these elements evolved behaviors resembling spiking or oscillatory neurons.
Similarly, \cite{yamaguti2015mathematical} used genetic algorithms to maximize bidirectional information transfer between subnetworks. 
Subsequently~\citep{yamaguti2021functional}, they applied evolutionary adaptation to reservoir computing networks, enabling differentiation of output units into categories that respond specifically to corresponding input categories. This transformation from random to structured networks showed some similarities to evolutionary changes in vertebrate brains, particularly the development from the reptilian medial pallium to the mammalian hippocampus and neocortex.
\cite{kanemura2024} investigated reservoir computing models using genetic algorithms to simultaneously maximize information transfer and minimize network maintenance costs, revealing that sparse networks with a medullary structure emerge under these constraints.

These studies demonstrate that artificial neural networks can develop functional structures similar to those observed in biological systems when constrained by information-theoretic measures, such as MI or transfer entropy, highlighting the importance of information-theoretic principles in the self-organization of functionally differentiated neural structures. 

However, estimating and optimizing MI between higher-dimensional variables has long been considered a difficult problem due to the curse of dimensionality. 
In response to this challenge, researchers have typically relied on either specialized models with tractable forms that allow gradient-based optimization of information-theoretic measures, or evolutionary approaches that avoid direct gradient computation altogether.

\cite{belghazi18mine} proposed MINE, a method that manages the estimation of MI for any differentiable model and enables the optimization of MI through backpropagation algorithms, harnessing the computational power of GPU-based deep learning frameworks.

In the present paper, we use MINE to estimate and minimize MI between neural subgroups in RNN models, investigating how functionally differentiated structures emerge under this information-theoretic constraint. Although most previous studies have focused on maximizing information metrics to enhance information transfer or representation quality~\citep{linsker1988,bell1995,tanaka2009,yamaguti2021functional}, our approach uses the \textit{minimization} of MI between neural subgroups as an organizing constraint. This constraint encourages subgroups to develop statistically independent activity patterns, thereby promoting functional specialization. To our knowledge, this is the first study to systematically investigate MI minimization as a mechanism for inducing functional differentiation in the context of computational neuroscience modeling. We acknowledge that alternative mechanisms—such as unsupervised learning rules, sparse connectivity constraints, or task structure alone~\citep{yang2019task}—can also lead to functional specialization, and comparative studies of these different approaches would be valuable for future research.

\section{Models and Methods}

\subsection{Neural network architectures}

\begin{figure}
\centering
\includegraphics[width=\linewidth]{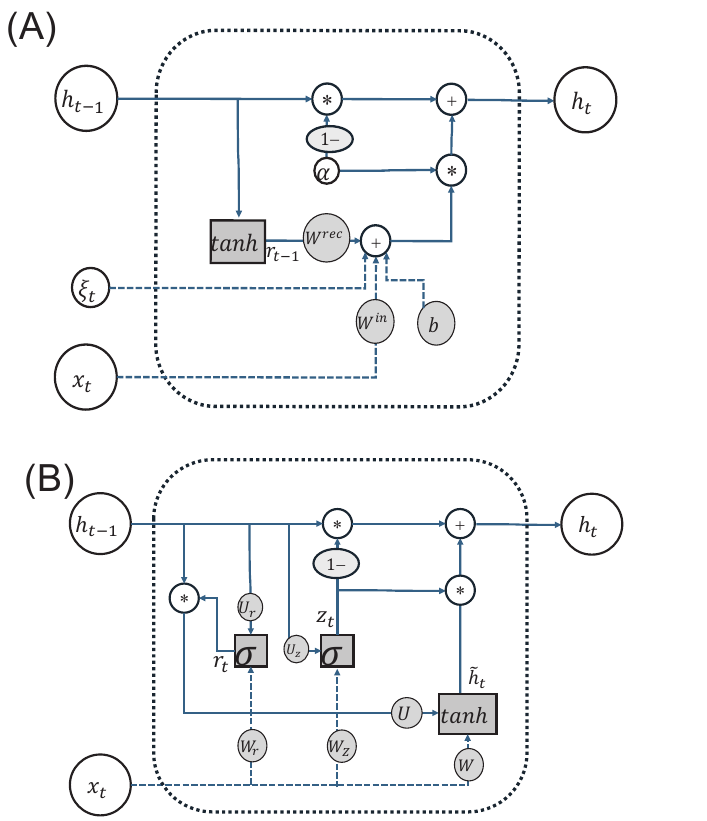}
\caption{Schematic of the hidden state update in the two RNN architectures used in this study. (A) Leaky-integrator RNN and (B) GRU. $\sigma$ is the sigmoid activation function and $\tanh$ is the hyperbolic tangent activation function. Weight parameters to be learned are represented by shaded circles}
\label{fig:model_update}
\end{figure}

We used two types of RNNs for the two experiments described in later sections: a simple leaky-integrator RNN and a gated recurrent unit (GRU) network~\citep{chung2014empirical}.
The leaky-integrator RNN~(Fig.~\ref{fig:model_update}A) was used for the working memory task, whereas the GRU~(Fig.~\ref{fig:model_update}B) was used for the chaotic signal separation task.
In both experiments, we incorporated the MINE algorithm for both estimation and optimization of MI between neural subgroups. 
Using these distinct network architectures and task paradigms allows us to evaluate the generality and robustness of our approach; obtaining consistent patterns of functional differentiation across these different conditions would strengthen the validity of our hypothesized principles.
In this section, we briefly describe these neural network architectures and the MINE algorithm used in our experiments. 

\subsubsection{Simple leaky-integrator RNN}\label{sec:simple_rnn}
For the working memory task, we used a simple leaky-integrator RNN, which incorporates a mechanism that allows the network to maintain past information while letting the information  gradually fade over time, a feature widely used in computational neuroscience models to approximate the firing rate dynamics of biological neurons better~\citep{sussillo2009,sussillo2014neural, song2016training,yang2019task}

The dynamics of a leaky-integrator RNN with $N$ units are given by
\begin{align}\label{eq:rnn}
h_t^{i} &= (1-\alpha)h_{t-1}^{i} + \alpha\Big(\sum_{j=1}^{N} w_{ij}^{\mathrm{rec}}r_{t-1}^{j} \nonumber \\
&+ \sum_{k=1}^{N_{\text{in}}} w_{ik}^{\mathrm{in}}x_t^k + b^i + \xi_{t}^{i}\Big),
\end{align}
where $h_t^i \in \mathbb{R}$ is the state of neuron $i$ at time $t$, 
$r_t^i = \tanh(h_t^i) \in \mathbb{R}$ is the firing rate of the neuron, $x_t^k$ is the $k$-th input signal at time $t$, $\xi_t^i \sim \mathcal{N}(0, 0.1)$ is independent Gaussian noise for neuron $i$,
$w^{\textrm{rec}}_{ij}$ is the $(i,j)$-th element of recurrent weight matrix $W^{\textrm{rec}} \in \mathbb{R}^{N \times N}$, 
$w^{\textrm{in}}_{ik}$ is the $(i,k)$-th element of input matrix $W^{\textrm{in}} \in \mathbb{R}^{N \times N_{\textrm{in}}}$, $b^i$ is the bias for neuron $i$, and $\alpha = 0.1$ is the leakage parameter controlling the decay rate.

The model outputs are given by the weighted sum of the recurrent layer output as
\begin{equation}\label{eq:output}
y_t^{j} = \sum_{i=1}^{N} w_{ji}^{\mathrm{out}}a_t^{i} + b_j^{\mathrm{out}},
\end{equation}
where $y_t^j \in \mathbb{R}$ is the $j$-th output of the model at time $t$,
$w_{ji}^{\mathrm{out}}$ is the $(j,i)$-th element of output weight matrix $W^{\mathrm{out}} \in \mathbb{R}^{N_{\text{out}} \times N}$, $b_j^{\mathrm{out}}$ is the bias for output $j$,
$a_t^i$ is the $i$-th element of the recurrent layer output (where $a_t^i = r_t^i$ for leaky-integrator RNNs), and $N_{\text{out}}$ is the number of output units.

\subsubsection{GRU} 
GRUs~\citep{chung2014empirical} are an extension of the RNN and process sequential data with a long sequence dependency. 
Although GRUs were proposed as a simplified alternative to the more complex long short-term memory model~\citep{hochreiter1997long}, they show comparable performance for various tasks.

GRUs have update gates and reset gates, which adaptively control memorization and forgetting process, thereby enabling them to retain more past information than traditional RNNs. 
The dynamics of GRUs with $N$ units are given by

\begin{align}
\begin{split}
    h_t^i &= \left(1-z_t^i\right)h_{t-1}^i+z_t^i{\widetilde{h}}_t^i \\
    z_t^i &= {\sigma(W_zx_t+U_zh_{t-1})}^i \\
    {\widetilde{h}}_t^i &= \tanh{{(Wx_t+U\left(r_t\odot h_{t-1}\right))}^i} \\
    r_t^i &= {\sigma(W_r x_t+U_rh_{t-1})}^i
    \label{eq:GRU}
\end{split}
\end{align}
where $t$ is time, $\sigma$ is the sigmoid function, 
$\odot$ is the Hadamard product of matrices, $z_t\in\left[0,1\right]^N$ is an update gate that controls how much of the new information to incorporate, $r_t\in\left[0,1\right]^N$ is a reset gate that determines which parts of the previous state to forget, ${\widetilde{h}}_t\in \mathbb{R}^N$ is candidate activation that represents new information that could be stored, and $h_t\in \mathbb{R}^N$ is the output of hidden states, hereafter simply called the hidden state.
$W_z$, $W$, and $W_{r}$ are weight matrices that determine how input $x_t \in \mathbb{R}^{N_{\textrm{in}}}$ affects variables $z_t$, $\widetilde{h}_t$, and $r_t$, respectively, and $U_{z}$, $U$, and $U_{r}$ are matrices that determine how the previous internal state affects these variables.
Upper right subscript $i$ denotes the $i$-th element of the vector.
The GRU output is also given by the weighted sum of the hidden states, $h_t$, as in Eq.~\eqref{eq:output}, where $a_t^i = h_t^i$ for GRU networks.

\subsection{MINE} 

MINE is a method developed by \cite{belghazi18mine} that uses deep neural networks to construct a differentiable estimator for MI. 
The differentiable nature of this estimator enables the backpropagation of MI gradients, allowing for neural network parameters to be optimized with respect to information-theoretic objectives by incorporating the estimated MI value into the loss function for training the model.
In our study, this capability is crucial because it allows the MI to be minimized between different neural subgroups while simultaneously optimizing task performance.

MI is an information-theoretic measure of the statistical dependence between two random variables, quantifying the amount of information shared between them. 
MI between random variables $X$ and $Y$ can be defined as the reduction in uncertainty about $X$ when $Y$ is known, as 
\begin{equation}
I\left(X;Y\right)=H\left(X\right)-H\left(X\middle| Y\right),
\end{equation}
where $H$ is the Shannon entropy and $H(X|Y)$ is the conditional Shannon entropy of variable $X$
given knowledge of variable $Y$.

MI can be expressed equivalently in terms of the Kullback--Leibler (KL) divergence as
\begin{equation} \label{eq:mi_kl} 
I\left(X;Y\right)=D_{KL}\left(\mathbb{P}_{XY}\parallel\mathbb{P}_X\otimes\mathbb{P}_{Y}\right), 
\end{equation}
where $\mathbb{P}_X$ and $\mathbb{P}_{Y}$ are marginal distributions of $X$ and $Y$, respectively, $\mathbb{P}_{XY}$ is their joint distribution, and $\mathbb{P}_X\otimes\mathbb{P}_{Y}$ is the product of the marginal distributions. This product distribution has the same marginals as $\mathbb{P}_{XY}$ but assumes that $X$ and $Y$ are statistically independent.

KL-divergence $D_{KL}$ between probability distributions $\mathbb{P}$ and $\mathbb{Q}$
is defined as
\begin{equation}
D_{KL}\left(\mathbb{P}\parallel\mathbb{Q}\right)=\mathbb{E}_\mathbb{P}\left[\log{\frac{d\mathbb{P}}{d\mathbb{Q}}}\right],
\end{equation}
where $\mathbb{P}$ and $\mathbb{Q}$ are distributions on compact domain $\Omega$. 
The KL-divergence of $\mathbb{P}$ relative to $\mathbb{Q}$ can be defined when $\mathbb{P}$ is absolutely continuous with respect to $\mathbb{Q}$.

Following \cite{donsker1983}, the KL-divergence can be expressed as a dual representation,
\begin{equation} 
D_{KL}\left(\mathbb{P}\parallel\mathbb{Q}\right)=\sup_{T:\Omega\to \mathbb{R}}{\mathbb{E}_\mathbb{P}\left[T\right]-\log(\mathbb{E}_\mathbb{Q}\left[e^T\right])},
\end{equation}
where the supremum is taken over all functions $T$ such that the two expectations are finite.

Applying this dual representation to MI by setting $\mathbb{P} = \mathbb{P}_{XY}$ and $\mathbb{Q} = \mathbb{P}_X \otimes \mathbb{P}_Y$ from Eq.~\eqref{eq:mi_kl}, we obtain
\begin{equation}
I(X;Y) = \sup_{T:\Omega\to \mathbb{R}}{\mathbb{E}_{\mathbb{P}_{XY}}\left[T\right]-\log(\mathbb{E}_{\mathbb{P}_X \otimes \mathbb{P}_Y}\left[e^T\right])}.
\end{equation}

The key insight of MINE is to approximate this supremum using function $T_\theta: \Omega \to \mathbb{R}$ parameterized by a neural network with parameters $\theta \in \Theta$, 
where $\Theta$ is the set of all possible parameters.
The neural network is trained using gradient-based optimization to maximize this objective, effectively approximating the supremum. The resulting estimate of MI is given by
\begin{equation} 
\widehat{{I(X;Y)}}={\sup}_{\theta\in\mathrm{\Theta}}\mathbb{E}_{\mathbb{P}_{XY}}\left[T_\theta\right]-\log(\mathbb{E}_{\mathbb{P}_X\otimes{\mathbb{P}}_Y}\left[e^{T_\theta}\right]),
\label{eq:mine}
\end{equation}
where $\widehat{{I(X;Y)}}$ is the estimated MI, which provides a lower bound on the true MI.

\subsection{Experimental setup}\label{sec:models_and_methods}

This subsection details our computational framework for investigating how functional differentiation emerges under information-theoretic constraints. 

Our framework consists of two interconnected components: a main model (MM) and a submodel (SM). 
The MM is an RNN that develops functional differentiation under MI constraints and is trained to perform specific tasks while its internal structure evolves under constraints. 
The SM is a neural network estimator that quantifies the MI between designated neuronal subgroups within the MM. 
Crucially, the SM estimates MI and provides gradients that allow MI to be minimized between these subgroups during training, thereby encouraging functional specialization.

In the following subsections, we detail the specific architectures, training procedures, and analytical methods used to evaluate the emergence of structural and functional modules in the trained networks.

\subsubsection{Main model}

The main model (MM) implements the RNN architectures that perform our experimental tasks while developing functional differentiation under MI constraints. 
Different RNN architectures are used for the two tasks, as described below.

For the working memory task, we use the leaky-integrator RNN represented by Eqs.~\eqref{eq:rnn} and \eqref{eq:output}.
For the chaotic signal separation task, we use a GRU network represented by Eqs.~\eqref{eq:GRU} and \eqref{eq:output}. 

A key aspect of our approach is the division of recurrent layer neurons into two equal-sized groups, $g^{(1)}=\{1, ..., N/2\}$ and $g^{(2)}=\{(1+N/2), ..., N\}$, with MI minimization between these groups driving functional differentiation during training.

In both tasks, the MM produces two output vectors, $y^{(1)}_t$ and $y^{(2)}_t$ (Fig.~\ref{fig:tasks}). The dimension, $M$, of each output vector is 1 for the working memory task and 3 for the signal separation task. Thus, the total output dimension is $N_{\text{out}} = 2M$, decomposed into two separate $M$-dimensional outputs.

These two outputs are calculated as weighted sums of the recurrent layer output,
\begin{equation}  
y^{(j)}_t = W^{(j)}_{y} a_t + b^{(j)}_{y},
\label{eq:output_task}
\end{equation}
where $j \in \{1,2\}$ indexes the output vectors, $a_t$ is the output of the recurrent layer at time $t$, $W^{(j)}_{y}$ is an $(M \times N)$ weight matrix connecting the recurrent layer to output $j$, and $b^{(j)}_{y}$ is the corresponding $M$-dimensional bias vector.
The definition of $a_t$ depends on the network architecture of $a_t = r_t$ for the working memory task (leaky-integrator RNN) and $a_t = h_t$ for the chaotic signal separation task (GRU).

\subsubsection{Working memory task}\label{sec:wm_task}

We use a 2-bit working memory task that requires the network to maintain two independent binary states simultaneously~(Fig.~\ref{fig:tasks}A). The task structure is as follows.

\begin{enumerate}
    \item The network receives four distinct input channels: two ON channels (ON$_1$, ON$_2$) and two OFF channels (OFF$_1$, OFF$_2$), where the subscripts indicate which memory bit (1 or 2) the channel affects.
    
    \item When a brief pulse (of amplitude $1.0$) is presented on the ON channel for a memory bit, the corresponding output should switch to and maintain the ON state (target value of 1.0).
    
    \item When a brief pulse is presented on the OFF channel for a memory bit, the corresponding output should switch to and maintain the OFF state (target value of $-1.0$).
    
    \item Between pulses, all input channels remain at zero, and the network must maintain the current state of both memory bits without additional input.
\end{enumerate}

The network has two output units ($N_{\text{out}}=2$) that correspond to the two memory bits. Each output is trained to maintain the appropriate state ($1.0$ for ON or $-1.0$ for OFF) based on the most recent pulse received, with pulses affecting one memory bit not disturbing the other.
The input pulses are generated randomly with a probability of 0.002 per time step per channel and width of 40 time steps. Training sequences are 2,000 time steps with an initial transient period of 1,000 time steps.

Task loss $L_{\textrm{task}}$ is defined as the mean squared error between the network's outputs and the target values,
\begin{equation}
L_{\textrm{task}} = \frac{1}{2L} \sum_{t=1}^{L} \sum_{j=1}^{2} (y_t^{(j)} - \hat{y}_t^{(j)})^2,
\end{equation}
where $y_t^{(j)}$ is the network's output for memory bit $j$ at time $t$, $\hat{y}_t^{(j)}$ is the corresponding target value, and $L$ is the sequence length.

\subsubsection{Chaotic signal separation task}\label{sec:sp_task}

 In this task, the MM receives a mixed input signal $x_t= x_t^{(1)}+x_t^{(2)}$, where $x_{t}^{(1)}$ and $x_{t}^{(2)}$ are the 3-dimensional signals generated by the Lorenz and R\"ossler systems, respectively, as described in the Appendix.
Thus, the input to MM is a single 3-dimensional time series representing the superposition of these two chaotic systems~(Fig. \ref{fig:tasks}B).

The MM generates two output time series, $y_t^{(1)}$ and $y_t^{(2)}$, each of which is a 3-dimensional signal that should approximate the original Lorenz and R\"ossler signals, respectively. 

As with the working memory task,
the MM undergoes supervised training, where task loss $L_{\textrm{task}}$ is the mean squared error between the outputs and their respective target signals,
\begin{equation}
L_{\textrm{task}} = \frac{1}{2L} \sum_{t=1}^{L} \left( \|y_t^{(1)} - x_t^{(1)}\|^2 + \|y_t^{(2)} - x_t^{(2)}\|^2 \right),
\end{equation}
where $L$ is the sequence length. During both training and evaluation, the input and target signals are generated by numerically solving differential equations Eqs. (\ref{eq:Lorenz}) and (\ref{eq:Rossler}) with random initial values.

After training, the task performance is evaluated by calculating the coefficient of determination $R^2$ for each variable in each output signal as
\begin{equation}
R^2 = 1-\frac{\sum_{t=1}^{L}\left(s_t-\hat{s}_t\right)^2}{\sum_{t=1}^{L}\left(s_t-\bar{s}\right)^2},
\end{equation}
where $s_t$ is a target signal component (either a component of $x_t^{(1)}$ or $x_t^{(2)}$), $\hat{s}_t$ is the corresponding network output component from either $y_t^{(1)}$ or $y_t^{(2)}$, and $\bar{s}$ is the mean of the target signal component. This metric quantifies the proportion of variance in the target signal that is captured by the network's output, with values closer to 1.0 indicating better performance.

\subsubsection{Sub model}\label{sec:submodel}

The sub model (SM) is a feed-forward neural network represented by $T_\theta$, where $\theta$ is the network parameters. 
The SM serves two key functions: estimating the MI between two neuronal groups in the MM and providing gradients to minimize MI during training.
The SM focuses on two neural subgroups from the recurrent layer in the MM,
$h_t^{(1)}= (h_1, ..., h_{N/2})$ and $h_t^{(2)}= (h_{N/2+1},...,h_N)$, which correspond to the neuron index sets, $g^{(1)}=\{1, ..., N/2\}$ and $g^{(2)}=\{(N/2+1), ..., N\}$, respectively. 
The input to the SM consists of these two vectors, each of length $N/2$. The SM architecture is a feed-forward neural network that produces a scalar value. Details of the SM architecture are provided in the Appendix.

To estimate the MI between the two neural groups indexed by $g^{(1)}$ and $g^{(2)}$, we use the following steps.
\begin{enumerate}
   \item  Input signals are fed into the MM to generate a time series of paired internal states $(h_t^{(1)},h_t^{(2)})$. 
  \item  To improve the learning stability, Gaussian noise with zero mean and standard deviation $\sigma_{\text{noise}}$ is independently added to each element of these vectors. 
  \item We collect these pairs across time steps and across samples in mini-batches, creating a dataset of sample size $(n_{\textrm{batch}} \times L)$ for the SM, where $L$ is the length of the input signal and $n_{\textrm{batch}}$ is the batch size.
  \item Following the MINE approach (Eq.~\eqref{eq:mine}), we compute the first term of the MI estimator by feeding the original paired samples, $(h_t^{(1)}, h_t^{(2)})$, to $T_\theta$ and calculating the average output as
  \begin{equation}
   \frac{1}{n_{\textrm{batch}}  L} \sum_{b=1}^{n_{\textrm{batch}}} \sum_{t=1}^{L} T_\theta(h_t^{(1,b)}, h_t^{(2,b)}),
  \end{equation}
  where superscript $b$ indicates the sample index in the mini-batch.
 \item  For the second term, we create independent samples by shuffling the pairs within the dataset, approximating samples from the product of marginal distributions. We then calculate
 \begin{equation}
  \log\left(\frac{1}{n_{\textrm{batch}}  L} \sum_{b=1}^{n_{\textrm{batch}}} \sum_{t=1}^{L} e^{T_\theta(h_t^{(1,b)}, h_{u(b,t)}^{(2,c(b,t))})}\right)
 \end{equation}
 where $c(b,t)$ and $u(b,t)$ are the indices of the randomly permuted samples and time steps, respectively.
  \item The estimated MI is the difference between these two terms, following Eq.~\eqref{eq:mine}.
   
\end{enumerate}

\subsubsection{L2 regularization}

L2 regularization is applied to encourage energetically economical network structures in our model.
 This method adds a penalty term to the loss function based on the squared magnitude of the weight parameters of
\begin{equation}
L_{\textrm{reg}} = \frac{1}{2} \sum_{w \in \mathcal{W}} w^2,
\end{equation}
where $\mathcal{W}$ is the set of all weight parameters in the MM. 

The inclusion of L2 regularization is important for our study of functional differentiation, which is systematically investigated in the section on chaotic signal separation task results. 

\subsubsection{Adversarial training process}

\begin{figure}[tbhp]
        \centering 
        \includegraphics[width=\linewidth]{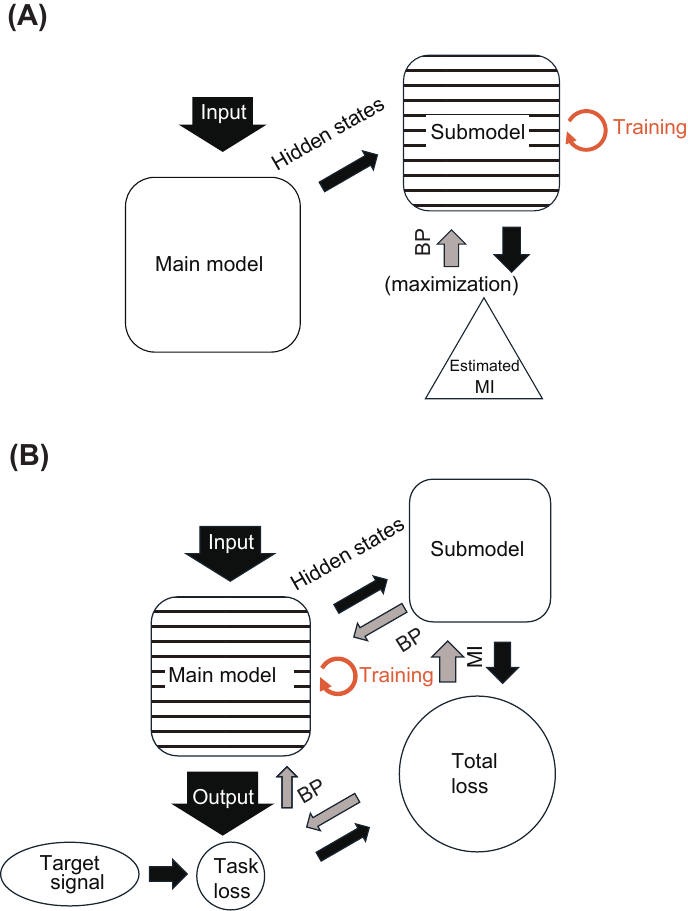}
        \caption{Training process of the MM and SM. 
        Training phases A and B are repeated alternately.
        BP indicates backpropagation. Horizontal hatching patterns  indicate which network is being optimized. (A) SM training. Maximizes MI estimate between neural subgroups using MINE. (B) MM training. Minimizes total loss including task performance and MI between the two neural subgroups} 

    \label{fig:trainings}
\end{figure}

The training process involves alternating optimization between the SM and the MM (Fig.~\ref{fig:trainings} A and B).
The process uses an adversarial relationship with respect to MI, where the SM learns to estimate MI better by maximizing it, as represented by Eq.~\eqref{eq:mine}, whereas the MM learns to perform the task and minimize the MI. 
This creates a minimax game similar to generative adversarial networks~\citep{goodfellow2014}, but focused on MI optimization rather than data generation.

To ensure stable training dynamics, we implement mini-batch training in which multiple input sequences are processed simultaneously. For the working memory task, each mini-batch contains $n_{\text{batch}}$ randomly generated pulse sequences. For the chaotic signal separation task, each mini-batch consists of $n_{\text{batch}}$ pairs of Lorenz and R\"ossler time series with randomly sampled initial conditions. 

Initially, the SM undergoes 100 training iterations to develop a reasonable MI estimator before beginning MM training.
Subsequently, we adopt an alternating schedule where the MM is trained once for every 20 training iterations of the SM.
This asymmetric schedule ensures that the SM maintains accurate MI estimation throughout training.

The training procedure implements an adversarial relationship between the SM and MM.
First, the SM's parameters, $\theta$, are updated to maximize the estimated MI (Fig.~\ref{fig:trainings}A). Input signals from the current mini-batch are fed into the MM to generate sequences of paired internal states $(h_t^{(1)}, h_t^{(2)})$ for the two neural subgroups. Then, the SM processes these paired states to estimate the MI between the two groups using the MINE algorithm (Eq. \eqref{eq:mine}). Finally, the SM's parameters are updated using a stochastic gradient descent optimizer to maximize the estimated MI.
 
 Second, for training the MM (Fig.~\ref{fig:trainings}B), 
input signals were fed into by the RNN to produce sequences of paired internal states and task outputs.  Then, the total loss for the MM is calculated as
\begin{equation}
L = L_{\textrm{task}} + \lambda_{\textrm{reg}} L_{\textrm{reg}} + \lambda_I \hat{I},
\end{equation}
where
$\hat{I}$ is the MI estimated by the SM,
$\lambda_{\textrm{reg}}$ is the weight for the regularization term, and
$\lambda_I$ is the weight for the MI minimization.
The MM's parameters are updated using the optimizer to minimize this total loss.
The inclusion of the MI term in the MM's loss function drives the formation of functionally differentiated modules because neural groups develop to minimize their statistical dependencies.

\subsubsection{Modularity and separability}  

We use modularity as a quantitative measure of the extent to which a network can be divided into distinct modules or communities. 
A higher modularity value indicates that connections within modules are dense whereas connections between modules are sparse~\citep{newman2018networks}.
In our analysis, the group membership of each neuron is predetermined and remains fixed throughout training, as specified in Section MM. 
The modularity index Q measures how well the learned network structure aligns with this predefined partition into groups $g^{(1)}$ and $g^{(2)}$. 
Following the approach by \cite{newman2004analysis} for weighted networks, we define modularity as follows. 
Consider a undirected, weighted, and connected network with $N$ nodes divided into $C$ different groups, where group membership of node $i$ is denoted by $s_i$. Let $A_{ij} \geq 0$ represent the weight of the connection between nodes $i$ and $j$, $k_i = \sum_j A_{ij}$ be the sum of all connection weights to node $i$, and $A_{\text{total}} = \frac{1}{2}\sum_{ij} A_{ij}$ be the total sum of all connection weights in the network. 
Modularity $Q$ is then defined as
\begin{equation}
Q = \frac{1}{2A_{\text{total}}}\sum_{ij}{(A_{ij}-\frac{k_{i}k_{j}}{2A_{\text{total}}})\delta(s_i,s_j)},
\end{equation}
where $\delta(s_i,s_j)$ is the Kronecker delta function, equal to 1 if nodes $i$ and $j$ belong to the same group and 0 otherwise. This measure quantifies how much the actual connection pattern deviates from what would be expected in a random network with the same node degrees.

In our analysis, we calculate modularity for two types of matrices to distinguish functional and structural organization as follows.

\textit{(1) Functional modularity} ($Q_{\textrm{cor}}$): Calculated from the correlation matrix between the activities of neurons in the recurrent layer, this measure quantifies the extent to which the dynamics of the network exhibit modular organization. This approach is analogous to functional connectivity analysis commonly used in neuroscience~\citep{friston2011functional}.

\textit{(2) Structural modularity} ($Q_{\textrm{str}}$): Calculated from the recurrent weight matrix of the recurrent layer, this measure quantifies the extent to which the physical connections in the network form distinct modules.

To compare the two matrices, we first symmetrize the recurrent weight matrix by taking absolute values and adding it to its transpose, $A^{\textrm{sym}} = |W| + |W^T|$, where $W$ is the original weight matrix.
However, the correlation matrix is already symmetric by definition, only requiring the absolute value operation to ensure non-negativity.
From both matrices, diagonal terms are removed to focus on the interactions among distinct nodes.

To quantify the degree of functional and structural separation among neural groups, we define three separability measures that assess how distinctly the two groups process inputs and generate outputs. These separability indices ($D$-measures) differ from modularity indices ($Q$-measures) in that they specifically evaluate the relationship among neural groups and their input/output connections, rather than measuring general community structure within the network.

Output separability $D_{\text{out}}$ measures the degree to which each neural group preferentially connects to one of the two output channels. This metric quantifies the structural bias in output connections as
\begin{equation}
D_{\text{out}} = \left| \frac{ \sum_{i=1}^{N}\sum_{k=1}^{M} (-1)^{s_{i}} \left( \left| w_{y,ki}^{(2)} \right| - \left| w_{y,ki}^{(1)} \right| \right) }{ \sum_{i=1}^{N}\sum_{k=1}^{M} \left( \left| w_{y,ki}^{(2)} \right| + \left| w_{y,ki}^{(1)} \right| \right) } \right|
\label{eq:output_separability}
\end{equation}
where $s_i \in \{1,2\}$ denotes the group membership of neuron $i$ and $w_{y,ki}^{(j)}$ represents the $(k,i)$-th element of weight matrix $W^{(j)}_y$ connecting the recurrent layer to output $j$ (Eq.~\eqref{eq:output_task}). Values approaching 1 indicate strong output specialization, whereas values near 0 suggest equal contribution to both outputs.

The input separability ($D_{\text{in}}$) is defined only for the working memory task, where input channels can be grouped naturally according to their target memory bits. This measure quantifies how distinctly each neural group responds to inputs associated with different memory bits as
\begin{equation}
D_{\text{in}} = \left| \frac{ \sum_{i=1}^{N} \sum_{k=1}^{2} (-1)^{s_i} \left( \left| w^{\text{in}1}_{ik} \right| - \left| w^{\text{in}2}_{ik} \right| \right) }{ \sum_{i=1}^{N} \sum_{k=1}^{2} \left( \left| w^{\text{in}1}_{ik} \right| + \left| w^{\text{in}2}_{ik} \right| \right) } \right|,
\label{eq:input_separability}
\end{equation}
where $w^{\text{in}j}_{ik}$ is the input weight from the $k$-th channel of the $j$-th input group (corresponding to memory bit $j$) to neuron $i$. Indices $k \in \{1,2\}$ correspond to the ON and OFF channels, respectively, for each memory bit.

The correlation separability ($D_{\text{cor}}$) measures functional specialization by quantifying how differently each neural group correlates with the two output channels. Unlike the weight-based measures above, this metric captures the actual functional relationships during network operation as
\begin{equation}
D_{\text{cor}} = \left| \frac{ \sum_{i=1}^{N} (-1)^{s_i} \left( \left| c_{i}^{(2)} \right| - \left| c_{i}^{(1)} \right| \right) }{ \sum_{i=1}^{N} \left( \left| c_{i}^{(2)} \right| + \left| c_{i}^{(1)} \right| \right) } \right|,
\label{eq:correlation_separability}
\end{equation}
where 
\begin{equation}
c_i^{(j)} = \frac{1}{M}\sum_{k=1}^{M} \left| \text{corr}(h^i_t, y^{(j)}_{t,k})\right|
\label{eq:correlation_output}
\end{equation}
represents the average correlation coefficient between the activity of neuron $i$ and all elements of output vector $y^{(j)}_t$.

\section{Working memory task results}\label{sec:wm_results}

\subsection{Task performance}

\begin{figure}[ht]
        \centering
        \includegraphics[width=\linewidth]{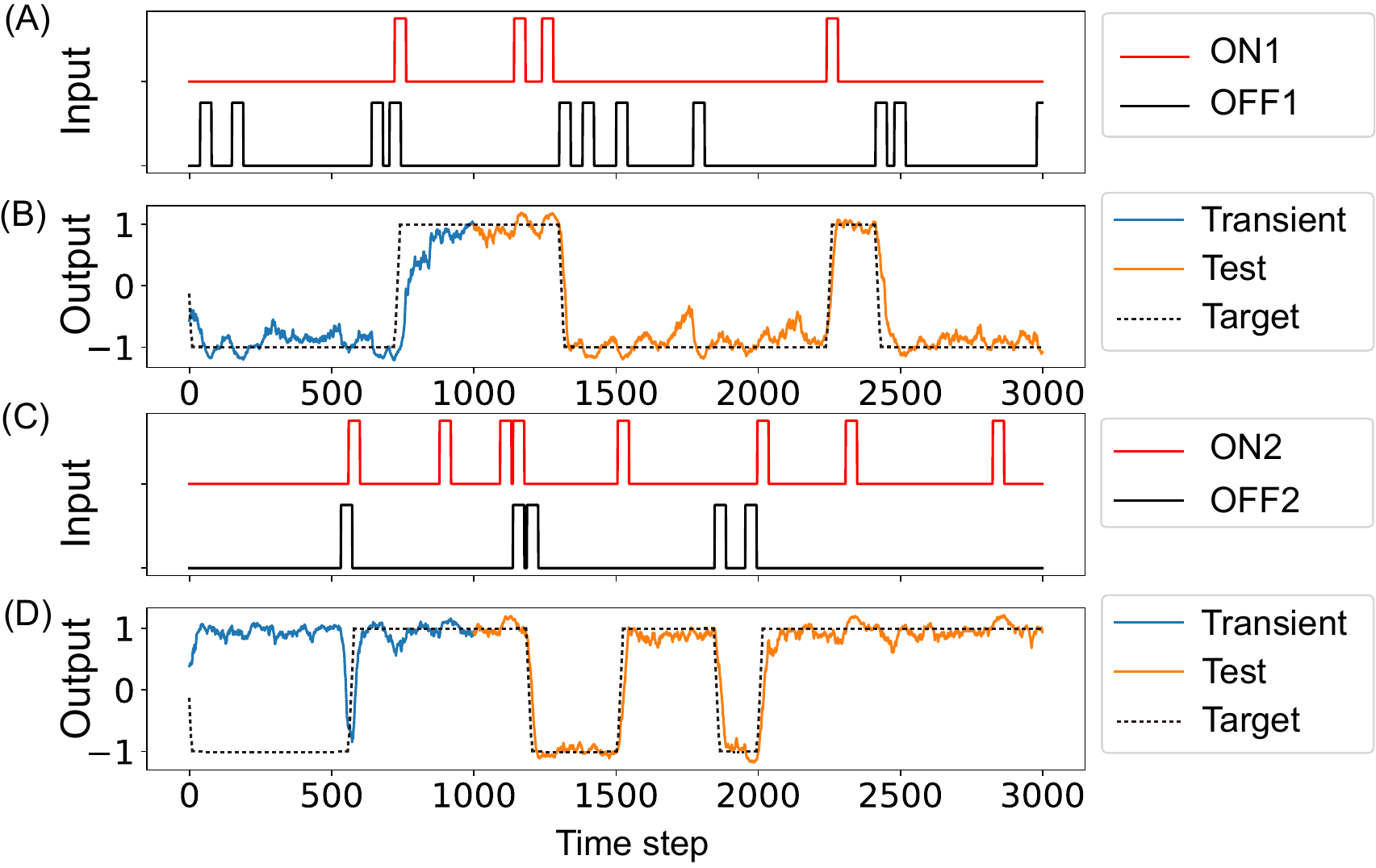}
        \caption{Dynamics of the working memory task. The four panels, from top to bottom, show: (A) input pulses for $\text{ON}_1$  and $\text{OFF}_{1}$ signals, (B) network output (solid) compared with target signal (dashed) for memory bit 1, (C) input pulses for $\text{ON}_{2}$ and $\text{OFF}_2$ signals, and (D) network output (solid) compared with target signal (dashed) for memory bit 2. In (B) and (D), the first 1,000 time steps are a transient period during which the outputs are not evaluated for the task loss calculation} 
        \label{fig:wm_dynamics}
\end{figure}

The network learned to perform the 2-bit working memory task, exhibiting the ability to maintain two independent memory bits across prolonged time intervals and respond appropriately to ON and OFF pulses. 
Figure \ref{fig:wm_dynamics} illustrates a representative example of the network's dynamics during the working memory task, showing the close alignment between the network's output and the target signals for both memory bits. 
The first 1,000 time steps are a transient period during which the network stabilizes its internal dynamics, and outputs during this period were excluded from task loss calculation to ensure fair performance evaluation.

To quantify task performance systematically, we defined success as achieving correct output states for more than 90\% of the time steps. 
Using this criterion, all 20 independent trials achieved successful performance when trained with MI minimization (100\% success rate). Networks trained without MI minimization also achieved successful performance (20/20 trials, 100\% success rate). These results indicate 
that the MI constraint does not impair task learning capability.

\subsection{Emergence of functional differentiation}

\subsubsection{Neuron--output correlation analysis}

\begin{figure}[ht]
        \centering
        \includegraphics[width=\linewidth]{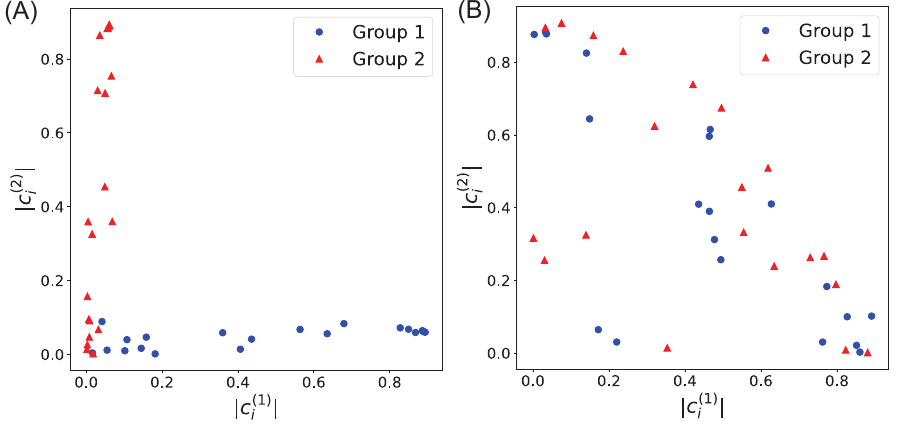}
        \caption{Correlation between neural activity and output signals in the working memory task. Each point represents a neuron, with coordinates determined by its absolute correlation with output signal 1 ($|c_i^{(1)}|$) and output signal 2 ($|c_i^{(2)}|$). (A) Results with MI minimization. (B) Results without MI minimization. Neurons in group 1 are represented by circles and those in group 2 by triangles} 
        \label{fig:wm_correlation}
\end{figure}

To analyze the emergence of functional differentiation in the network, we examined the correlation patterns between neural activities and output signals. Figure \ref{fig:wm_correlation} shows scatter plots of correlation coefficients between each neuron's activity and the two output signals, with coordinates determined by correlations with output signal 1 ($c_i^{(1)}$) and output signal 2 ($c_i^{(2)}$).

When trained with MI minimization (Fig.~\ref{fig:wm_correlation}A), neurons in group 1 (circles) and group 2 (triangles) formed distinct clusters near their respective axes, demonstrating clear functional specialization where each neural subgroup primarily maintained one of the two memory bits. 

In contrast, without MI minimization (Fig.~\ref{fig:wm_correlation}B), both groups were intermixed throughout the correlation space and exhibited mixed selectivity where most neurons respond to both memory bits. 
This result is particularly notable because the four input channels (ON$_1$, OFF$_1$, ON$_2$, OFF$_2$) are statistically identical, providing no explicit cues about which channels should control which memory bit. 
Without the MI constraint, the network has no pressure to align the two predefined neural subgroups with the two memory bits, resulting in both groups contributing to both outputs. The clear separation observed with MI minimization (Fig.~\ref{fig:wm_correlation}A) therefore demonstrates that the information-theoretic constraint drives the network to discover an organization that aligns predefined groups with the task's latent structure.

Importantly, both organizational strategies achieved comparable task performance. Networks trained with MI minimization showed mean squared error of $0.146 \pm 0.0320$, while networks without MI minimization achieved $0.166 \pm 0.053$ ($p = 0.14$, two-sample $t$s-test; all trials used independent random initializations), indicating no significant performance difference. 
Furthermore, among networks trained with MI minimization, we found no significant correlation between the final estimated MI values and task loss (Pearson $r = 0.0046$, $p = 0.98$), suggesting that MI minimization primarily serves as an organizational constraint that shapes network structure rather than directly optimizing task performance. 

To quantify this functional specialization pattern, we calculated the correlation separability index $D_{\text{cor}}$ for both conditions. Networks trained with MI minimization achieved $D_{\text{cor}} = 0.623 \pm 0.316$
(mean $\pm$ standard deviation across 20 independent trials),
 indicating strong functional differentiation, whereas networks without MI minimization showed significantly lower separability ($D_{\text{cor}} = 0.072 \pm 0.057$).

\subsubsection{Correlation matrix analysis}

\begin{figure}[htb]
        \centering
        \includegraphics[width=\linewidth]{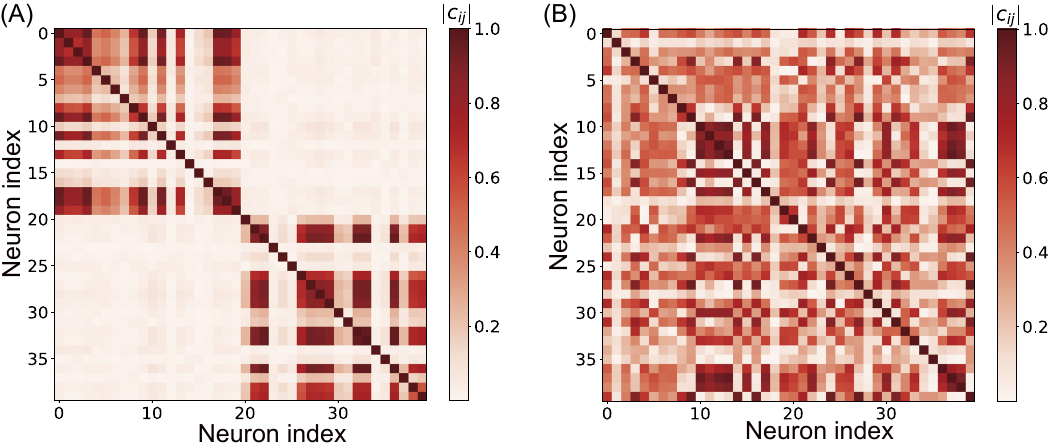} 
        \caption{Correlation matrix of neural activities in the working memory task. Neurons from groups $g^{(1)}$ and $g^{(2)}$ are arranged along both axes. (A) Results with MI minimization. (B) Results without MI minimization} 
        \label{fig:wm_corr_mat}
\end{figure}

We analyzed the functional organization further by examining the pairwise correlations between all neurons in the network. 
Figure \ref{fig:wm_corr_mat} shows the correlation matrix of neural activities, 
with neurons from groups $g^{(1)}$ and $g^{(2)}$ arranged along both axes. 
When trained with MI minimization (Fig.~\ref{fig:wm_corr_mat}A), the correlation matrix revealed a striking block-diagonal structure, with high positive correlations between neurons within the same group (two darker blocks along the diagonal) and weak or negative correlations between neurons in different groups.

This correlation structure indicates that neurons within each group exhibited similar activity patterns while functioning largely independently from neurons in the other group. 
Functional modularity index $Q_{\text{cor}}$ calculated from this correlation matrix reached a value of $0.299 \pm 0.153$ (mean $\pm$ standard deviation over 20 trials),
quantifying the strong functional segregation between the two neural subgroups. 
This high degree of functional modularity confirmed that MI minimization induced the network to develop functionally differentiated modules where each module specializes in processing different aspects of the task.
In contrast, no clear block-diagonal structure was observed when the same network was trained without MI minimization (Fig.~\ref{fig:wm_corr_mat}B). In this case, $Q_{\text{cor}}$ was $-0.0162 \pm 0.012$, indicating a lack of functional modularity.

\subsection{Structural connectivity analysis}

In addition to examining functional differentiation through correlation patterns, 
we analyzed how the structural connectivity of the network evolved during training with MI minimization.
This analysis focused on three aspects of structural connectivity: the recurrent connection weights among neurons within the RNN, the input weights from input channels to RNN neurons, and the output weights connecting the RNN to the output units.

\subsubsection{Recurrent weight differentiation} 

\begin{figure}[ht]
        \centering
        \includegraphics[width=\linewidth]{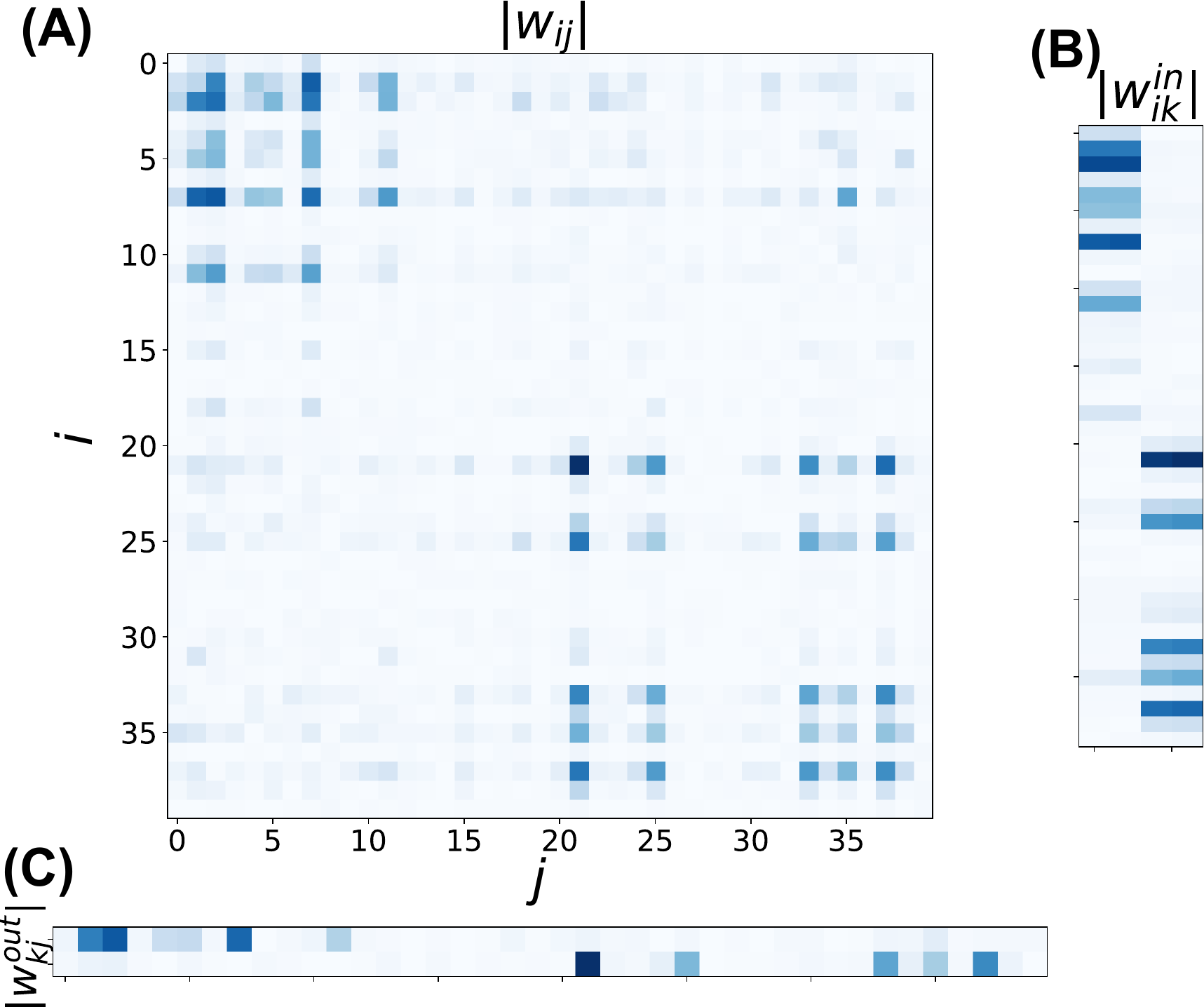}
        \caption{Weight matrix differentiation in the working memory task. (A) Recurrent weight matrix $W^{\text{rec}}$. (B) Input weight matrix $W^{\text{in}}$. (C) Output weight matrix $W^{\text{out}}$.
        In all panels, absolute values of weights are shown, with darker colors indicating stronger connections
        }
        \label{fig:wm_weights}
\end{figure}

Figure \ref{fig:wm_weights}A shows the absolute values of the recurrent weights after training, with neurons from both groups ordered along both axes. 
The recurrent weight matrix exhibited a modular structure, with stronger connections within groups (darker diagonal blocks) and weaker inter-group connections.

Across 20 independent training runs, the structural modularity index showed $Q_{\text{str}} = 0.114 \pm 0.049$ (mean $\pm$ standard deviation). Although approximately half the trials developed a clear modular structure, the remainder failed to exhibit pronounced structural organization, resulting in $Q_{\text{str}} < 0.1$. 
These cases appeared to represent local minima where functional differentiation emerged without corresponding structural reorganization (detailed analysis in the Failure modes of structural differentiation section). 

Despite this variability, the consistent emergence of functional modularity demonstrates that MI minimization effectively reduces information sharing between neural subgroups. 
Additionally, the modest $Q_{\text{str}}$ values compared with $Q_{\text{cor}}$ suggest that complete structural segregation is not required for functional specialization.

\subsubsection{Input weight differentiation} 

We analyzed input weight matrix $W^{\text{in}}$ to determine whether the neural subgroups became specialized for processing specific input channels. Figure \ref{fig:wm_weights}B depicts the input weight strengths, with neurons arranged along the vertical axis and the four input channels (ON$_1$, OFF$_1$, ON$_2$, and OFF$_2$) along the horizontal axis.

Neurons in group $g^{(1)}$ developed stronger connections from input channels related to the first memory bit (ON$_1$ and OFF$_1$), whereas neurons in group $g^{(2)}$ developed stronger connections from channels related to the second memory bit (ON$_2$ and OFF$_2$). 
Input separation index $D_{\text{in}}$ reached a value of $0.623 \pm 0.31$, indicating strong specialization in input processing.

\subsubsection{Output weight differentiation} 

Finally, we examined output weight matrix $W^{\text{out}}$ to determine whether the functionally differentiated neural subgroups developed specialized output projections. Figure \ref{fig:wm_weights}C shows the absolute values of output weights, with neurons from groups $g^{(1)}$ and $g^{(2)}$ arranged along the horizontal axis and the two output units along the vertical axis.

A clear division of specialization emerged in the output connectivity: neurons in group $g^{(1)}$ formed strong projections to output unit 1 (responsible for the first memory bit) and weak projections to output unit 2, whereas neurons in group $g^{(2)}$ showed the opposite pattern. 
This structural specialization was quantified using output separation index $D_{\text{out}}$, which reached a value of $0.526 \pm 0.261$ by the end of training, indicating a high degree of output weight specialization.

\subsection{Failure modes of structural differentiation}
\label{sec:wm_failure_modes}

Although MI minimization consistently induced functional differentiation, structural modularity emergence was more variable. In 10 out of 20 training runs, the recurrent weight matrix failed to develop a clear modular structure ($Q_{\text{str}} < 0.1$), despite achieving functional specialization. 

\begin{figure}[tbh]
        \centering
        \includegraphics[width=\linewidth]{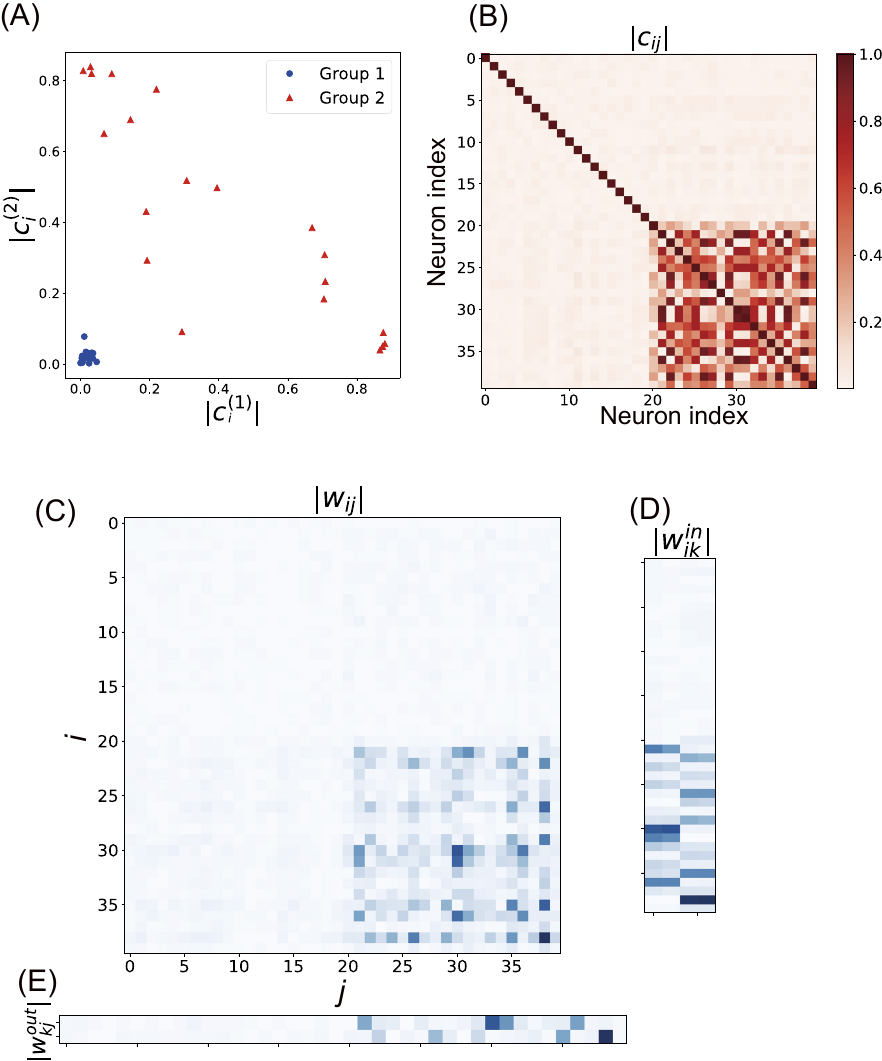}
        \caption{Analysis of a failed trial in the working memory task. (A) 
        Correlation between neurons and output signals. (B)
        Correlation matrix. (C-E) Weight matrices 
        }
        \label{fig:wm_failure_modes} 
\end{figure}

Figure \ref{fig:wm_failure_modes} shows a representative failed case.  The weight matrices (Fig.~\ref{fig:wm_failure_modes}C-E) reveal an asymmetric solution where group 2 neurons dominated both input processing and output generation, whereas group 1 connections were largely eliminated. The neuron--output correlation analysis (Fig.~\ref{fig:wm_failure_modes}A) shows that although group 2 neurons exhibited stronger correlations with both of the two output signals, group 1 neurons had weak or negligible correlations. 
Furthermore, the correlation matrix (Fig.~\ref{fig:wm_failure_modes}B) shows that group 1 neurons were not well-correlated with each other, indicating that they were not functioning as a coherent module. Because of noise,
group 1 neurons exhibited higher random activity, leading to reduced MI between the groups.

This asymmetric pattern suggests that MI minimization can be satisfied through functional silencing of one group rather than balanced specialization. 
These cases represent suboptimal local minima where the optimization constraints are satisfied without achieving the intended modular architecture.

\subsection{Time course of modularity development}

A key finding in our study is that functional differentiation precedes structural differentiation during training. We analyzed the 10 trials (50\%) that achieved robust structural differentiation ($Q_{\text{cor}} > 0.1$) to examine this temporal relationship.

\begin{figure}[ht]
        \centering 
        \includegraphics[width=\linewidth]{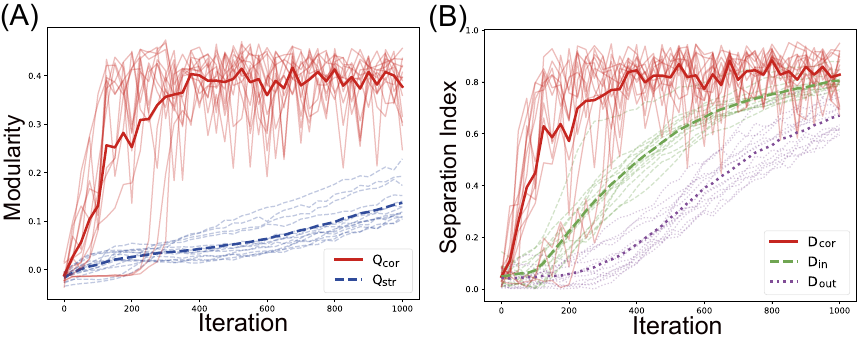}
        \caption{Time course of modularity development in the working memory task. (A) Evolution of the functional modularity index ($Q_{\text{cor}}$, solid line) and structural modularity index ($Q_{\text{str}}$, dashed line). (B) Evolution of the correlation-based separation index ($D_{\text{cor}}$, solid line), input weight separation index ($D_{\text{in}}$, dashed line), and output weight separation index ($D_{\text{out}}$, dotted line). Bold lines represent mean values across 10 successful trials} 
        \label{fig:modularity_development}
\end{figure}

Figure \ref{fig:modularity_development} reveals distinct temporal dynamics. Functional modularity ($Q_{\text{cor}}$) rises rapidly within the first 300 training steps, whereas structural modularity ($Q_{\text{str}}$) increases more gradually. Similarly, correlation-based separation ($D_{\text{cor}}$) develops faster than both the input and output weight separation indices.

This temporal precedence was consistent across different initializations, indicating that neural activity patterns adapt first to task requirements, establishing functional specialization before significant synaptic reorganization occurs.

\section{Chaotic signal separation task results}
\label{sec:sp_results}

\subsection{Task performance}

\begin{figure}[ht]
        \centering
        \includegraphics[width=\linewidth]{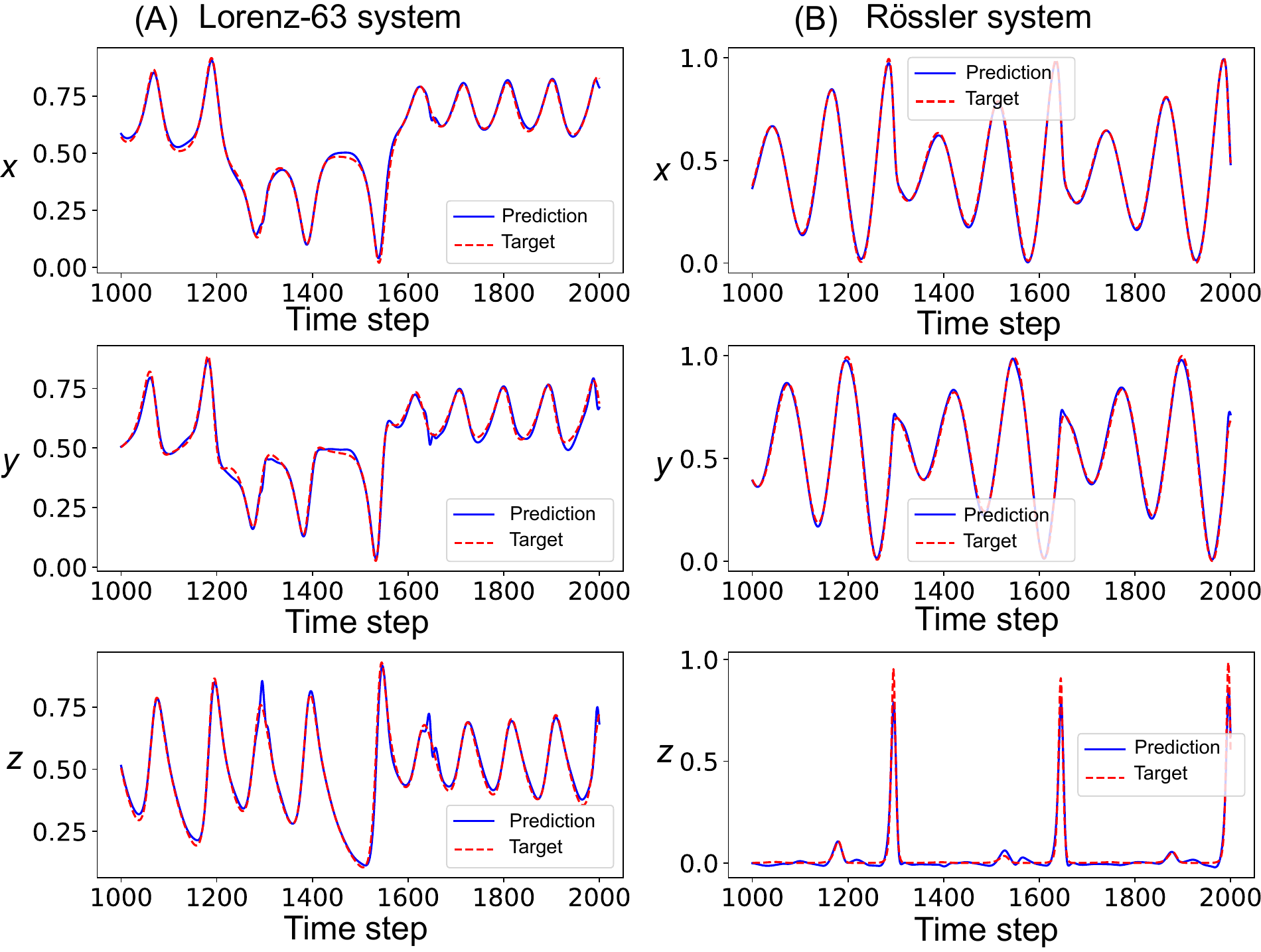}
        \caption{Signal separation performance for chaotic time series. (A) Separation of Lorenz system components $(x, y, z)$. (B) R\"ossler system components. Network outputs are shown as solid blue lines and target signals are shown as dashed red lines } 
        \label{fig:sp_re_separation}
\end{figure}

The GRU network learned to separate the mixed chaotic signals, achieving high reconstruction accuracy for both the Lorenz and R\"ossler time series. Figure \ref{fig:sp_re_separation} shows representative examples of the network's performance.

Quantitative evaluation using the coefficient of determination ($R^2$) showed excellent performance across all variables.
The average $R^2$ value across all six output variables (three for the Lorenz system and three for the R\"ossler system) was $0.973 \pm 0.013$ (mean $\pm$ standard deviation over 20 trials).
This high accuracy demonstrates that the network learned the complex nonlinear dynamics required for signal separation while maintaining the constraint of MI minimization.

\subsection{Functional connectivity analysis}

To evaluate whether MI minimization induced functional specialization in the network, we analyzed activity correlations following the same approach used for the working memory task.

\subsubsection{Neuron--output correlation analysis}
We examined the correlation patterns between individual neurons' activities and the network outputs to characterize functional specialization. For each neuron $i$ in the GRU layer, we calculated the mean absolute correlation coefficients with each output system following Eq.~\eqref{eq:correlation_output}.

\begin{figure}[ht]
        \centering
        \includegraphics[width=\linewidth]{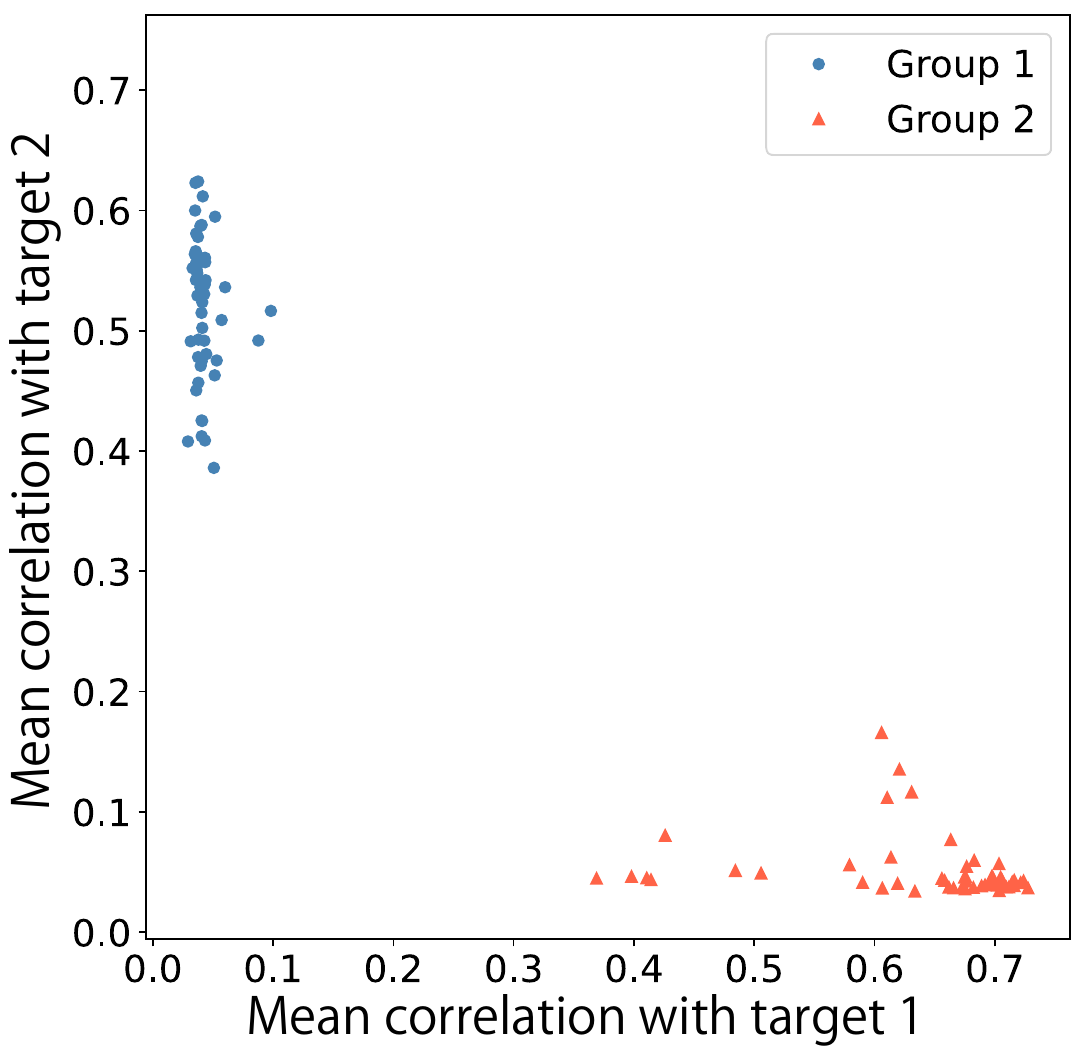}
        \caption{Correlation between GRU neuron activities and output signals. Each point represents a neuron, with coordinates determined by its mean correlation with the Lorenz reconstruction output (horizontal axis) and R\"ossler reconstruction output (vertical axis). Circles represent neurons from group $g^{(1)}$, and triangles represent neurons from group $g^{(2)}$} 
        \label{fig:sp_scatter_plot}
\end{figure}

Figure \ref{fig:sp_scatter_plot} shows a scatter plot of these correlation values, with each point representing a single neuron. 
Neurons in group 1 (circles) cluster near the vertical axis, indicating strong correlation with the R\"ossler output and weak correlation with the Lorenz output. Conversely, neurons in group 2 (triangles) cluster near the horizontal axis, showing the opposite pattern.

This clear separation demonstrates functional specialization comparable to the working memory task, indicating strong functional differentiation between the neural subgroups.
In this case, the average of the correlation separability index, $D_{\text{cor}}$, reached $0.494 \pm 0.285$ (mean $\pm$ standard deviation over 20 trials).
The variance among trials is higher than in the working memory task because the network occasionally failed to develop clear specialization where each neural module corresponds perfectly to all three components $(x, y, z)$ of either the Lorenz or R\"ossler systems. In some cases, the 3 + 3 dimensional output mapping did not align neatly with the 3/3 division between the two neural groups, leading to lower $D_{\text{cor}}$ values.

\subsubsection{Correlation matrix analysis}

\begin{figure}[tbh]
        \centering
        \includegraphics[width=0.95\linewidth]{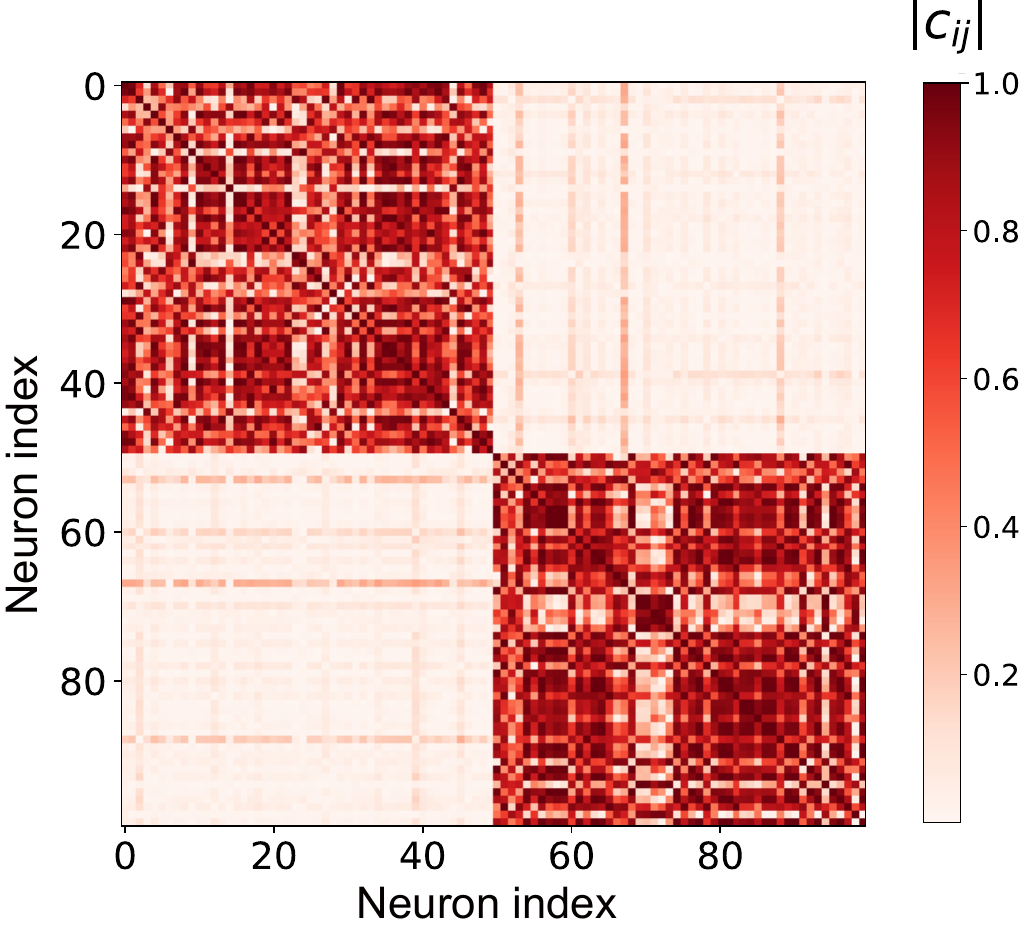}
        \caption{ Correlation matrix of neural activities in the signal separation task. 
        Neurons from groups $g^{(1)}$ and $g^{(2)}$ are arranged along both axes}
        \label{fig:sp_corr_mat} 
\end{figure}

We analyzed the correlation structure between neural activities to assess functional modularity. 
Figure \ref{fig:sp_corr_mat} displays the correlation matrix between all pairs of neurons in the GRU layer.

When trained with MI minimization, the correlation matrix exhibited a pronounced block-diagonal structure, with high positive correlations among neurons within the same group (two darker blocks along the diagonal) and weak or negative correlations among neurons in different groups.
The functional modularity index calculated from this matrix was $Q_{\text{cor}} = 0.375 \pm 0.090$, 
confirming that MI minimization effectively induced the emergence of functionally differentiated modules in the network.

\subsection{Structural connectivity analysis}

\subsubsection{Output weight specialization}

\begin{figure}[ht]
        \centering
        \includegraphics[width=\linewidth]{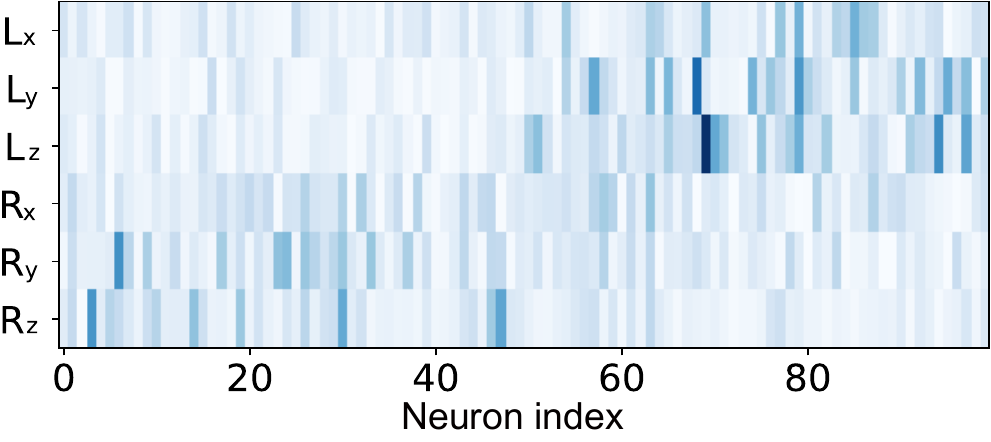}
        \caption{Output weight specialization in the chaotic signal separation task. The heatmap shows the absolute values of output weights connecting the GRU neurons (horizontal axis) to the two output targets (vertical axis). A distinguishable specialization is evident with group $g^{(1)}$ neurons (left) mainly connecting to the R\"ossler output and group $g^{(2)}$ neurons (right) mainly connecting to the Lorenz output} 
        \label{fig:sep_output_weights}
\end{figure}

Figure \ref{fig:sep_output_weights} shows the absolute values of the output weight matrix, with neurons from groups $g^{(1)}$ and $g^{(2)}$ arranged along the horizontal axis and the six output targets, consisting of Lorenz and R\"ossler components, along the vertical axis. 

In this example, a visible pattern of specialization emerged: neurons in group $g^{(1)}$ (left half) developed strong projections to the R\"ossler reconstruction output while maintaining weak projections to the Lorenz reconstruction output, with neurons in group $g^{(2)}$ (right half) showing the opposite pattern.

This specialization was quantified using the output separation index, $D_{\text{out}}$ (Eq.~\eqref{eq:output_separability}). 
The trained network achieved $D_{\text{out}} = 0.173 \pm 0.151$, 
compared with $D_{\text{out}} = 0.021 \pm 0.015$ for networks trained without MI minimization, indicating moderate but significant ($p<0.0001$, two-sample $t$-test; all trials used independent random initializations) output specialization.
The variance in $D_{\text{out}}$ across trials is also higher than in the working memory task, reflecting incomplete specialization in some cases, as described in the Neuron--output correlation analysis section. 

\subsubsection{Recurrent weight modularity}
\begin{figure}[ht]
        \centering
        \includegraphics[width=\linewidth]{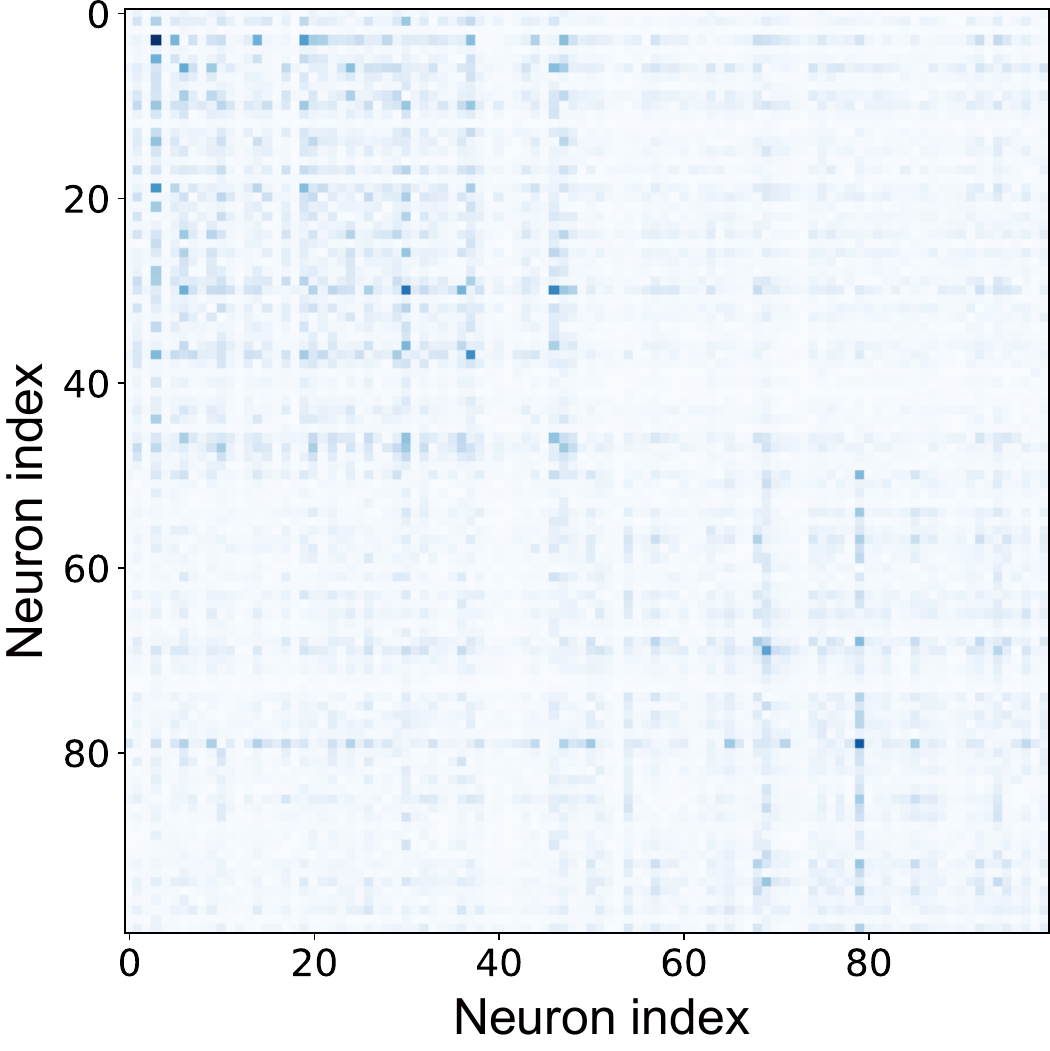}  
        \caption{Recurrent weight modularity in the chaotic signal separation task. The heatmap shows the symmetrized recurrent weight matrix with neurons from groups $g^{(1)}$ and $g^{(2)}$ arranged along both axes. The block-diagonal structure indicates stronger connections within groups (darker diagonal blocks) compared with weaker connections among groups} 
        \label{fig:sp_recurrent_weights}
\end{figure}

Figure \ref{fig:sp_recurrent_weights} shows the symmetrized recurrent weight matrix, $U_{\text{sym}} = |U| + |U|^T$, where $U$ is the original recurrent weight matrix from the GRU (Eq.~\eqref{eq:GRU}). 

The matrix revealed a modular structure with subtly stronger connections within groups (slightly darker blocks along the diagonal) compared with connections between groups.

The structural modularity was quantified using modularity index $Q_{\text{str}}$, which achieved a value of $0.088 \pm 0.035$. 
Although this value was lower than the functional modularity observed in correlation patterns, 
it was significantly higher than that for the networks trained without MI minimization ($Q_{\text{str}} = -0.0048 \pm 0.0056$; $p<0.0001$, $t$-test),
demonstrating that MI minimization influences both the functional dynamics and the underlying structural connectivity of the network.

\subsection{Community detection validation} 

To validate the correspondence of the observed modularity with the predefined groups, $\{g^{(1)}, g^{(2)}\}$, we used community detection algorithms to identify the optimal partition that maximizes modularity without prior knowledge of group assignments.

Using the fast greedy algorithm \citep{clauset2004finding}, we identified the community structure that maximizes modularity in the correlation matrix. The similarity between this data-driven partition and our predefined groups was evaluated using normalized MI as
\begin{equation}
I_n = \frac{I(G; C)}{\frac{1}{2}[H(G) + H(C)]}
\end{equation}
where $G$ is the predefined partition $\{g^{(1)}, g^{(2)}\}$, $C$ is the community detection result, $I(G; C)$ is the MI between the two partitions, and $H(\cdot)$ is entropy~\citep{newman2018networks}.

Specifically, for each neuron $i$, we record its community membership in both partitions: $g_k$ for the predefined partition ($k \in \{1,2\}$) and $c_j$ for the detected partition ($j \in \{1,2,...\}$). From the set of tuples $(i, c_j, g_k)$ for all $N$ neurons, we compute empirical probabilities by counting: $P(c_j) = n_{c_j}/N$, $P(g_k) = n_{g_k}/N$, and $P(c_j, g_k) = n_{c_j,g_k}/N$, where $n_{c_j}$, $n_{g_k}$, and $n_{c_j,g_k}$ are the corresponding neuron counts. 
The entropy $H(C)$, $H(G)$, and mutual information $I(G;C)$ are then calculated from these probabilities using their standard definitions for discrete variables. The normalized MI ranges from 0 (independent partitions) to 1 (identical partitions).

For networks trained with MINE, the normalized MI was $I_n = 0.926 \pm 0.224$, indicating almost perfect agreement between the optimal community structure and our predefined groups. 
In contrast, networks trained without MINE achieved only $I_n = 0.018 \pm 0.014$, demonstrating that MI minimization is crucial for establishing the modular organization.

\begin{figure}[ht]
        \centering 
        \includegraphics[width=1.0\linewidth]{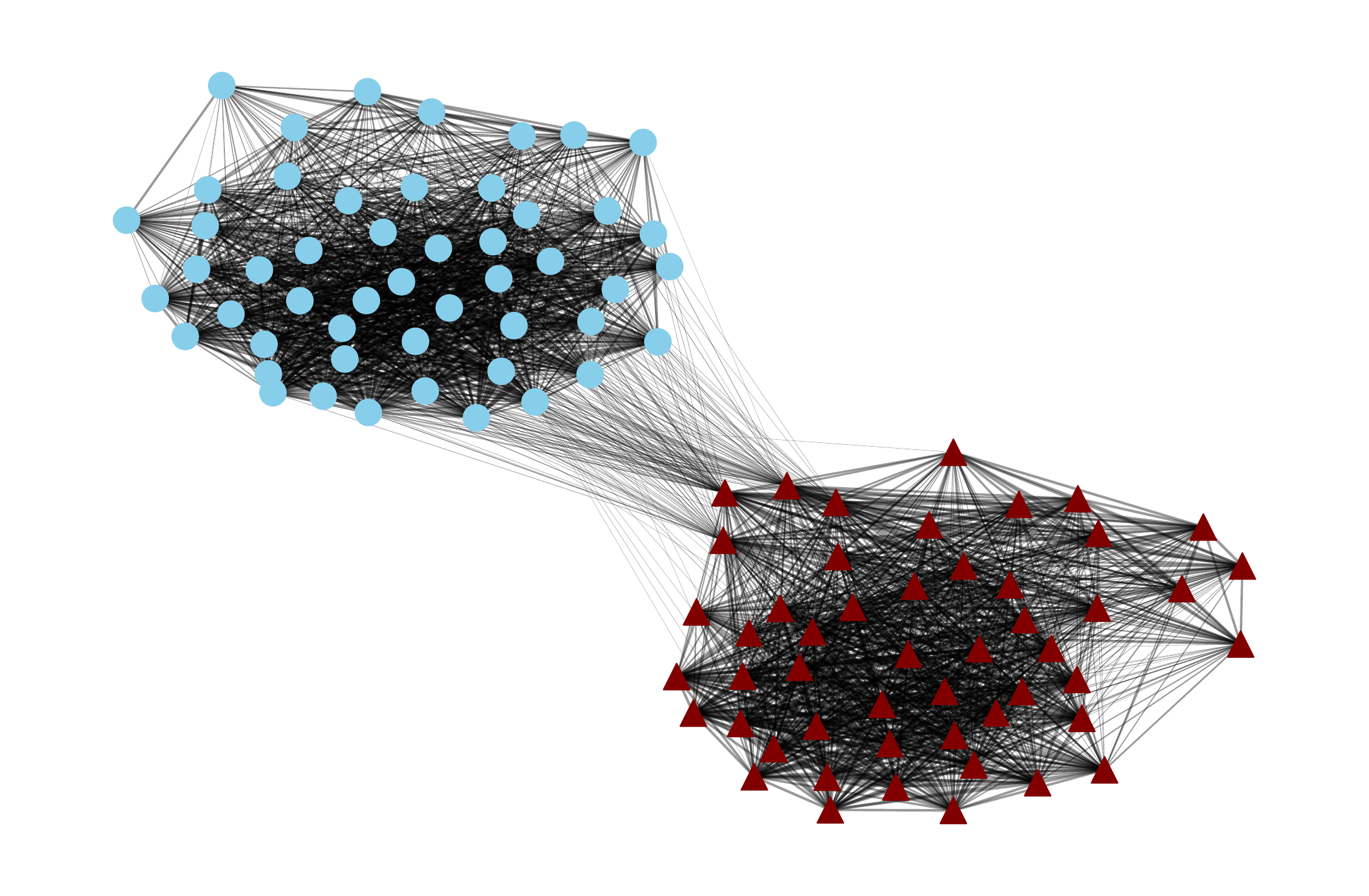}
        \caption{Community structure validation for networks trained with MI minimization. 
        Node shapes indicate which group the neuron belonged to (triangles: $g^{(1)}$, circles: $g^{(2)}$), and the node color (light blue or dark red) indicates the community membership estimated by the community detection algorithm. In this case, the community detection algorithm identified the two groups, with all neurons in $g^{(1)}$ assigned to one community and all neurons in $g^{(2)}$ assigned to another. 
        Edges are drawn to indicate correlations, with only the top 50\% strongest correlations shown for clarity 
        }
        \label{fig:sp_module_structure}
\end{figure}

Figure \ref{fig:sp_module_structure} illustrates this validation graphically, showing that the network consisted of two strongly-connected modules. 
The network layout was generated using a spring-based force-directed algorithm, where nodes with stronger correlations were positioned closer together. 
This validation confirmed that MI minimization increased modularity values and specifically drove the formation of the intended functional architecture, rather than arbitrary clustering patterns.

\subsection{Effect of L2 regularization on network structure}

Our analysis revealed that L2 regularization is crucial in determining whether functional differentiation translates into structural reorganization. 
We compared networks trained with and without L2 regularization to isolate its effects on modular structure formation.

With L2 regularization, networks achieved significant output separation ($D_{\text{out}}=0.173 \pm 0.151$) and structural modularity ($Q_{\text{str}}=0.088 \pm 0.035$), as shown in the Structural connectivity analysis section. 
 In contrast, networks trained without L2 regularization showed weaker structural organization ($D_{\text{out}} = 0.041 \pm 0.032$, $Q_{\text{str}} = 0.013 \pm 0.006$).
Notably, functional modularity and correlation separability remained relatively high, even without L2 regularization ($Q_{\text{cor}} = 0.380 \pm 0.075$ and $D_{\text{cor}} = 0.460 \pm 0.285$), suggesting that MI minimization alone is sufficient to induce functional specialization. 

This pattern suggests that MI minimization drives functional organization through activity patterns, whereas the translation into structural reorganization requires additional sparsity constraints. L2 regularization encourages sparse connectivity patterns that reinforce the functionally established modules, bridging the gap between functional and structural differentiation.

\subsection{Temporal development of modularity}

\begin{figure}[tbhp]
        \centering
        \includegraphics[width=\linewidth]{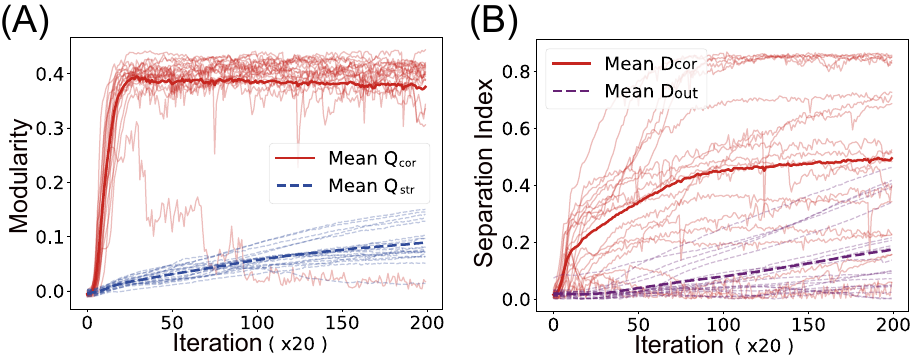}
        \caption{Time course of modularity and separability development in the chaotic signal separation task. (A) Development of functional modularity $Q_{\text{cor}}$ (solid line)  and  structural modularity $Q_{\text{str}}$ (dashed line) over training iterations. 
        (B) Development of correlation-based separation $D_{\text{cor}}$ (solid line) and output weight separation $D_{\text{out}}$ (dotted line).
        In both panels, bold lines represent mean values across 20 trials.
        The horizontal axis represents the number of training iterations, with one unit corresponding to 20 MM training cycles 
        }
        \label{fig:sp_modularity_time}
\end{figure}

Similar to the working memory task, functional differentiation preceded structural reorganization during training. Figure \ref{fig:sp_modularity_time}A tracks the evolution of both modularity indices.
Functional modularity $Q_{\text{cor}}$ rose rapidly in the earlier training iterations, whereas structural modularity $Q_{\text{str}}$ exhibited a more gradual increase.
The separation indices (Figure \ref{fig:sp_modularity_time}B) showed similar temporal patterns, with correlation-based separation $D_{\text{cor}}$ developing earlier than output weight separation $D_{\text{out}}$.
However, the variance across trials was larger than for modularity indices, reflecting the occasional difficulty in achieving complete specialization in separating the Lorenz and R\"ossler components.

Although the trial variance was larger than in the working memory task, the overall 
 temporal pattern mirrored the working memory task findings, confirming that the precedence of functional over structural organization is a general principle in networks trained with MI minimization constraints. The consistency across different tasks strengthens the hypothesis that information-theoretic constraints first shape neural activity patterns, which subsequently influence synaptic connectivity reorganization.
This finding has important implications for understanding how functional and structural brain networks might co-evolve during development and learning.

\section{Discussion}

This study demonstrates that minimizing MI among neural subgroups induces functionally differentiated structures in RNNs. Our key findings provide new insights into how information-theoretic constraints can drive neural organization.

\subsection{Main findings}

Networks trained with MI minimization achieved high task performance while developing distinct functional modules. Three key patterns emerged consistently. (1) Functional modularity exceeded structural modularity, indicating that activity-based organization is more pronounced than connectivity-based organization. (2) Functional modules emerged earlier than structural modules during training, with correlation patterns developing within earlier iterations compared with later iterations for weight reorganization. (3) L2 regularization increased robust structural modularity, suggesting that efficiency constraints are crucial in translating functional demands into structural organization.

It is important to clarify our use of the term "emergence" in describing functional differentiation. Our approach involves a top-down constraint (MI minimization) imposed at the system level, which might seem inconsistent with the notion of bottom-up emergence. However, we follow the framework of "self-organization with constraints" proposed by Tsuda et al.~\citep{tsuda2016,tsuda2022nature}, where system-wide constraints guide the spontaneous organization of internal components. In this framework, emergence refers to the process by which functional modules and specialized connectivity patterns arise through local synaptic modifications under global constraints, rather than being explicitly predetermined. 
The specific pattern of functional differentiation—which neurons specialize for which functions—is not specified by the constraint itself but emerges through the optimization process. This perspective aligns with developmental neuroscience, where genetic and activity-dependent constraints shape neural organization without fully specifying the final functional architecture~\citep{katz1996synaptic}.

\subsection{Biological relevance}

The observation that functional modularity exceeds structural modularity 
is interesting in relation to neurodevelopmental research, where the relationship between functional and structural connectivity remains an area of active investigation~\citep{friston1994functional,fujii1996dynamical,bullmore2009complex,hayashi2021}. 
The temporal precedence of functional over structural modularity observed in our networks aligns with neurodevelopmental processes where spontaneous activity patterns emerge before synaptic maturation~\citep{katz1996synaptic}. Our results suggest that information-theoretic constraints could provide organizing principles that shape early activity patterns, which subsequently guide structural refinement through activity-dependent mechanisms.

The requirement for efficiency constraints (modeled by L2 regularization) also has biological parallels. Metabolic pressures in the brain favor sparse connectivity patterns~\citep{laughlin2003communication}, and developmental pruning eliminates unnecessary connections while strengthening functional ones~\citep{huttenlocher1997regional}. Our findings indicate that both information optimization and efficiency constraints are necessary for complete functional differentiation.

 An important question is how MI minimization between neural subgroups might be implemented biologically. 
In our model, MI is estimated using an auxiliary network via MINE, which may not have a direct biological counterpart.
However, biologically plausible approximations may exist.
The computation of correlations between neural activities, which is central to MI estimation, is implemented through synaptic plasticity mechanisms such as anti-Hebbian learning~\citep{hebb1949} and
spatio-temporal learning rule, which was proposed by Tsukada \citep{tsukada1996,tsukada2005}.
These local learning rules can detect and modify statistical dependencies without explicit MI computation.
Second, the brain implements mechanisms for reducing statistical dependencies between representations, such as pattern separation~\citep{rolls2013} and sensory decorrelation~\citep{barlow1961}. While no specific brain region is known to explicitly compute MI between neural subgroups, the existence of correlation-based plasticity mechanisms suggests that biologically plausible approximations may be feasible.
Our approach can be viewed as implementing a top-down organizational constraint analogous to developmental signals that guide circuit formation~\citep{katz1996synaptic}.

\subsection{Computational implications}

From a machine-learning perspective, our method offers a principled approach for inducing modular architectures that maintain high task performance.
The ability to develop functionally specialized modules suggests applications in continual learning~\citep{veniat2021efficient} and multi-task scenarios~\citep{yang2019task}, where modules could be reused across related tasks or protected from catastrophic forgetting~\citep{serra2018overcoming}. 

The temporal development pattern we observed of functional followed by structural specialization suggests training strategies where initial phases focus on establishing functional specialization through information-theoretic constraints, followed by structural optimization through sparsity regularization.

\subsection{Limitations and future directions}

Our analysis focused on predefined two-group divisions with relatively simple tasks. 
For the working memory task, while the task has two independent memory channels, our results demonstrate that MI minimization is necessary to drive functional differentiation: without this constraint, networks develop mixed selectivity despite the task structure (Fig.~\ref{fig:wm_correlation}B).
Nevertheless, investigating MI minimization on tasks requiring stronger inter-module cooperation would provide additional evidence for the generality of our approach.

An important consideration is the relationship between modular organization and mixed selectivity. Our results show that both strategies achieved comparable task performance, and the absence of significant correlation between MI values and task performance confirms that MI minimization primarily serves as an organizational principle rather than a direct performance optimizer. This aligns with growing recognition in neuroscience that mixed selectivity—where individual neurons encode multiple task variables—can be advantageous for flexible computation and generalization~\citep{rigotti2013importance}.
The optimal balance between modular specialization and mixed selectivity likely depends on factors we did not systematically investigate, including robustness, generalization to novel task variants, and continual learning capabilities~\citep{serra2018overcoming,yang2019task}. Understanding when biological neural systems employ each organizational strategy, and how information-theoretic constraints interact with these principles, represents an important direction for future research.

Future research should investigate optimal group sizes using adaptive partitioning methods, test scalability on hierarchical tasks that are more complex, and systematically explore the relationship between MI minimization strength and differentiation degree. Additionally, extending our approach to spiking neural networks would strengthen biological relevance and enable direct comparisons with neuroscientific data.
A particularly promising direction involves extending our approach to multiple modules ($>2$). Current pairwise MI minimization between two groups could be generalized to minimize MI among $K$ neural subgroups simultaneously. This generalization would require developing efficient optimization algorithms that can handle the combinatorial complexity of multi-group MI estimation.
These extensions would enable the modeling of more realistic neural architectures with multiple specialized functional areas, better reflecting the hierarchical organization observed in biological neural systems~\citep{sporns2016book}. 

\subsection{Conclusions}

Our results demonstrate that information-theoretic constraints can serve as fundamental organizing principles in neural systems, driving the emergence of modular architectures that balance specialization with integration. The temporal precedence of functional over structural organization and the role of efficiency constraints provide insights into biological  neural development and the design of artificial neural architectures.


\newcounter{saveeqn}
\setcounter{saveeqn}{\value{equation}}

\backmatter

\begin{appendices}
\setcounter{equation}{\value{saveeqn}}
\section{Appendix}
\subsection{Chaotic signals}\label{sec:chaos}

For the signal separation task, we used two well-known chaotic dynamic systems as signal sources: the Lorenz-63 system~\citep{lorenz1963} and the R\"ossler system~\citep{rossler1976}. These systems were selected because they generate complex, non-periodic signals with distinct dynamic characteristics.

To generate input signals for our experiments, we numerically solved the differential equations for both systems and sampled 10,000 points after discarding the first 4,000 points to eliminate transient dynamics. The numerical solutions were sampled at discrete intervals of $\Delta t=0.00714$ for the Lorenz system and $\Delta t=0.05$ for the R\"ossler system.
These specific sampling intervals were chosen so that within 1,000 time steps of the RNN, we would observe approximately seven cycles of the basic oscillation of the R\"ossler system and eight cycles of the Lorenz system, providing sufficient temporal complexity for the task.

The Lorenz-63 system is given by the following differential equations.
\begin{equation}
        \left\{ \,
            \begin{aligned}
            & \frac{dx}{dt}=\sigma(y-x) \\
            & \frac{dy}{dt}=x\left(\rho-z\right)-y \\
            & \frac{dz}{dt}=xy-\beta z,
            \end{aligned}
        \right.
        \label{eq:Lorenz}
\end{equation}
Here, $x$, $y$, and $z$ are state variables and $\sigma$, $\rho$, and $\beta$ are control parameters. We used the standard parameter values $\sigma=10$, $\rho=28$, and $\beta=\frac{8}{3}$.

The R\"ossler system is given by
\begin{equation}
        \left\{ \,
            \begin{aligned}
            & \frac{dx}{dt}=-y-z \\
            & \frac{dy}{dt}=x+ay \\
            & \frac{dz}{dt}=b+xz-cz,
            \end{aligned}
        \right.
        \label{eq:Rossler}
\end{equation}
where $x$, $y$, and $z$ are state variables and $a$, $b$, and $c$ are control parameters.
We used the standard parameter values $a=0.2$, $b=0.2$, and $c=5.7$.

For our experiments, we combined these two chaotic signals by simple addition to create mixed input signals. 
The neural network was then trained to recover the original separate signals, a task that requires the network to differentiate and process the distinct dynamic patterns of each system.

\subsection{Parameters}\label{sec:param}
Here, we provide the parameters used in our two experiments.

\subsubsection{Working memory task}
The parameters used in the working memory task are as follows:
number of neurons, $N=40$; number of training iterations, $n_{\text{iter}}=1,000$; batch size, $n_{\text{batch}}=20$; and number of time steps during one training sequence, $T=1,000$. 
RMSProp optimizers with learning rates of $0.002$ for the RNN and $0.003$ for the MINE network were used.
For the MINE method, we used a neural network with two hidden layers, each containing $512$ neurons. The activation function for the hidden layers was ReLU, and the output layer used a linear activation function. The noise strength, $\sigma_{\text{noise}}$, added to the input to the SM was set to $0.1$. The number of training iterations for the SM in one cycle, $n_{sm}$, was set to 20.
The weight for MINE loss was $\lambda_I=0.1$ and that for L2 regularization was $\lambda_{\text{reg}} = 0.01$. The leakage parameter for the RNN was set to $\alpha = 0.1$.

\subsubsection{Chaotic signal separation task}
The parameters used in the chaotic signal separation task are as follows:
number of neurons, $N=100$; number of training iterations, $n_{\text{iter}}=4000$;
 batch size, $n_{\text{batch}}=50$; and number of time steps during one training sequence, $T = 10,000$. 
The number of training iterations for the SM in one cycle $n_{sm}$ was set to 20
Adam optimizers with learning rates of $0.001$ for the RNN and $0.01$ for the MINE network. 
For the MINE method, we used a neural network with three hidden layers, each containing 100 neurons. 
The activation function for the hidden layers was leaky ReLU with slope parameter $0.25$, and the output layer used a linear activation function. 
The noise strength, $\sigma_{\text{noise}}$, added to the input to the SM was set to $0.05$. 
 The weight for MINE loss was $\lambda_I$ = 0.005 and that for L2 regularization was $\lambda_{\text{reg}} = 0.0001$.

\end{appendices}

\section*{Statements and Declarations} 

\noindent\textbf{Competing Interests:} I. T. and Y. Y. are members of the editorial board of the journal. The authors have no other competing interests to declare.

\noindent\textbf{Funding:} This work was partly supported by JSPS KAKENHI Grant Number 23K11256 and JST CREST Grant Number JPMJCR17A4. Open access funding will be provided by Fukuoka Institute of Technology.

\noindent\textbf{Data Availability:} The code and data used in this study are available at \url{https://github.com/cncs-fit/mio_rnn}.

\noindent\textbf{Author Contributions:} Y. T. conceived and conducted the signal separation experiments, and wrote the original draft. I. T. conceptualized the study, supervised the project, and revised the manuscript. Y. Y. conceptualized the study, conceived and conducted the working memory experiments, supervised the project, and wrote the original and revised manuscript.
All authors reviewed and approved the final manuscript.


\begin{thebibliography}{55}
\providecommand{\natexlab}[1]{#1}
\providecommand{\url}[1]{{#1}}
\providecommand{\urlprefix}{URL }
\providecommand{\doi}[1]{\url{https://doi.org/#1}}
\providecommand{\eprint}[2][]{\url{#2}}
 \bibcommenthead


\bibitem[{Amari(1972)}]{amari1972}
Amari SI (1972) Learning patterns and pattern sequences by self-organizing nets of threshold elements. IEEE Trans Comput C-21(11):1197--1206. \doi{10.1109/T-C.1972.223477}

\bibitem[{Amari(1980)}]{amari1980}
Amari SI (1980) Topographic organization of nerve fields. Bull Math Biol 42(3):339--364. \doi{10.1016/S0092-8240(80)80055-3}

\bibitem[{Barlow(1961)}]{barlow1961}
Barlow HB (1961) Possible principles underlying the transformation of sensory messages. In: Rosenblith WA (ed) Sensory Communication. MIT Press, Cambridge, MA, pp 217--234


\bibitem[{Belghazi et~al.(2018)Belghazi, Baratin, Rajeshwar, Ozair, Bengio, Courville, and Hjelm}]{belghazi18mine}
Belghazi MI, Baratin A, Rajeshwar S, et~al (2018) Mutual information neural estimation. In: Dy J, Krause A (eds) Proceedings of the 35th International Conference on Machine Learning, Proceedings of Machine Learning Research, vol~80. PMLR, pp 531--540, \urlprefix\url{https://proceedings.mlr.press/v80/belghazi18a.html}

\bibitem[{Bell and Sejnowski(1995)}]{bell1995}
Bell AJ, Sejnowski TJ (1995) {An information-maximization approach to blind separation and blind deconvolution.} Neural Comput 7:1129--59. \doi{10.1162/neco.1995.7.6.1129}


\bibitem[{Brodmann(1909)}]{brodmann1909vergleichende}
Brodmann K (1909) Vergleichende Lokalisationslehre der Grosshirnrinde in ihren Prinzipien dargestellt auf Grund des Zellenbaues. Barth

\bibitem[{Bullmore and Sporns(2009)}]{bullmore2009complex}
Bullmore E, Sporns O (2009) Complex brain networks: graph theoretical analysis of structural and functional systems. Nat Rev Neurosci 10(3):186--198. \doi{10.1038/nrn2575}

\bibitem[{Caianiello(1961)}]{caianiello1961}
Caianiello ER (1961) Outline of a theory of thought-processes and thinking machines. J Theor Biol 1(2):204--235. \doi{10.1016/0022-5193(61)90046-7}

\bibitem[{Chung et~al.(2014)Chung, Gulcehre, Cho, and Bengio}]{chung2014empirical}
Chung J, Gulcehre C, Cho K, et~al (2014) Empirical evaluation of gated recurrent neural networks on sequence modeling. arXiv preprint arXiv:14123555 {\href{https://arxiv.org/abs/1412.3555}{{arXiv:1412.3555}}} {[cs.NE]}

\bibitem[{Clauset et~al.(2004)Clauset, Newman, and Moore}]{clauset2004finding}
Clauset A, Newman ME, Moore C (2004) Finding community structure in very large networks. Phys Rev E 70(6):066111. \doi{10.1103/PhysRevE.70.066111}

\bibitem[{Clune et~al.(2013)Clune, Mouret, and Lipson}]{clune2013}
Clune J, Mouret JB, Lipson H (2013) {The evolutionary origins of modularity.} Proc R Soc B 280(1755):20122863. \doi{10.1098/rspb.2012.2863}

\bibitem[{Donsker and Varadhan(1983)}]{donsker1983}
Donsker MD, Varadhan SRS (1983) Asymptotic evaluation of certain markov process expectations for large time. iv. Commun Pure Appl Math 36(2):183--212. \doi{10.1002/cpa.3160360204}

\bibitem[{Ellefsen et~al.(2015)Ellefsen, Mouret, and Clune}]{ellefsen2015}
Ellefsen KO, Mouret JB, Clune J (2015) Neural modularity helps organisms evolve to learn new skills without forgetting old skills. PLoS Comput Biol 11(4):1--24. \doi{10.1371/journal.pcbi.1004128}

\bibitem[{Elman(1990)}]{elman1990}
Elman J (1990) {Finding structure in time}. Cogn Sci 14(2):179--211. \doi{10.1207/s15516709cog1402_1}

\bibitem[{Espinosa-Soto and Wagner(2010)}]{espinosa-soto2010}
Espinosa-Soto C, Wagner A (2010) {Specialization can drive the evolution of modularity.} PLoS Comput Biol 6:e1000719. \doi{10.1371/journal.pcbi.1000719}

\bibitem[{Felleman and Van~Essen(1991)}]{felleman1991}
Felleman DJ, Van~Essen DC (1991) Distributed hierarchical processing in the primate cerebral cortex. Cereb Cortex 1(1):1--47. \doi{10.1093/cercor/1.1.1-a}

\bibitem[{Friston(1994)}]{friston1994functional}
Friston KJ (1994) Functional and effective connectivity in neuroimaging: a synthesis. Human brain mapping 2(1-2):56--78. \doi{10.1002/hbm.460020107}

\bibitem[{Friston(2011)}]{friston2011functional}
Friston KJ (2011) Functional and effective connectivity: A review. Brain Connect 1(1):13--36. \doi{10.1089/brain.2011.0008}

\bibitem[{Fujii et~al.(1996)Fujii, Ito, Aihara, Ichinose, and Tsukada}]{fujii1996dynamical}
Fujii H, Ito H, Aihara K, et~al (1996) Dynamical cell assembly hypothesis-theoretical possibility of spatio-temporal coding in the cortex. Neural Netw 9(8):1303--1350. \doi{10.1016/S0893-6080(96)00054-8}

\bibitem[{Glasser et~al.(2016)Glasser, Coalson, Robinson, Hacker, Harwell, and Yacoub}]{Glasser2016}
Glasser MF, Coalson TS, Robinson EC, et~al (2016) {A multi-modal parcellation of human cerebral cortex}. Nature 536(7615):171--178. \doi{10.1038/nature18933}

\bibitem[{Goodfellow et~al.(2014)Goodfellow, Pouget-Abadie, Mirza, Xu, Warde-Farley, Ozair, Courville, and Bengio}]{goodfellow2014}
Goodfellow IJ, Pouget-Abadie J, Mirza M, et~al (2014) Generative adversarial nets. In: Ghahramani Z, Welling M, Cortes C, et~al (eds) Advances in Neural Information Processing Systems, vol~27. Curran Associates, Inc., \urlprefix\url{https://proceedings.neurips.cc/paper_files/paper/2014/file/f033ed80deb0234979a61f95710dbe25-Paper.pdf}

\bibitem[{Hayashi et~al.(2021)Hayashi, Hou, Glasser, Autio, Knoblauch, Inoue-Murayama, Coalson, Yacoub, Smith, Kennedy, and {Van Essen}}]{hayashi2021}
Hayashi T, Hou Y, Glasser MF, et~al (2021) The nonhuman primate neuroimaging and neuroanatomy project. NeuroImage 229:117726. \doi{10.1016/j.neuroimage.2021.117726}

\bibitem[{Hebb(1949)}]{hebb1949}
Hebb DO (1949) The Organization of Behavior. Wiley, New York

\bibitem[{Hochreiter and Schmidhuber(1997)}]{hochreiter1997long}
Hochreiter S, Schmidhuber J (1997) Long short-term memory. Neural Comput 9(8):1735--1780. \doi{10.1162/neco.1997.9.8.1735}

\bibitem[{Hoerzer et~al.(2014)Hoerzer, Legenstein, and Maass}]{hoerzer_2014}
Hoerzer GM, Legenstein R, Maass W (2014) Emergence of {Complex} {Computational} {Structures} {From} {Chaotic} {Neural} {Networks} {Through} {Reward}-{Modulated} {Hebbian} {Learning}. Cereb Cortex 24(3):677--690. \doi{10.1093/cercor/bhs348}

\bibitem[{Hopfield(1982)}]{hopfield1982}
Hopfield JJ (1982) Neural networks and physical systems with emergent collective computational abilities. Proc Natl Acad Sci USA 79(8):2554--2558. \doi{10.1073/pnas.79.8.2554}


\bibitem[{Huttenlocher and Dabholkar(1997)}]{huttenlocher1997regional}
Huttenlocher PR, Dabholkar AS (1997) Regional differences in synaptogenesis in human cerebral cortex. J Comp Neurol 387(2):167--178. \doi{10.1002/(SICI)1096-9861(19971020)387:2<167::AID-CNE1>3.0.CO;2-Z}

\bibitem[{Ichikawa and Kaneko(2024)}]{ichikawa2024}
Ichikawa K, Kaneko K (2024) Bayesian inference is facilitated by modular neural networks with different time scales.  PLoS Comput Biol 20(3):1--21. \doi{10.1371/journal.pcbi.1011897}

\bibitem[{Kaneko(2006)}]{kaneko2006life}
Kaneko K (2006) Life: an introduction to complex systems biology. Springer

\bibitem[{Kaneko and Tsuda(2001)}]{kaneko2001book}
Kaneko K, Tsuda I (2001) Complex systems: chaos and beyond: a constructive approach with applications in life sciences. New York: Springer Verlag

\bibitem[{Kanemura and Kitano(2024)}]{kanemura2024}
Kanemura I, Kitano K (2024) Emergence of input selective recurrent dynamics via information transfer maximization. Sci Rep 14(1):13631. \doi{10.1038/s41598-024-64417-6}

\bibitem[{Kashtan and Alon(2005)}]{kashtan2005}
Kashtan N, Alon U (2005) {Spontaneous evolution of modularity and network motifs.} Proc Natl Acad Sci USA 102:13773--8. \doi{10.1073/pnas.0503610102}

\bibitem[{Katz and Shatz(1996)}]{katz1996synaptic}
Katz LC, Shatz CJ (1996) Synaptic activity and the construction of cortical circuits. Science 274(5290):1133--1138. \doi{10.1126/science.274.5290.1133}

\bibitem[{Kawai et~al.(2023)Kawai, Park, Tsuda, and Asada}]{kawai2023learning}
Kawai Y, Park J, Tsuda I, et~al (2023) Learning long-term motor timing/patterns on an orthogonal basis in random neural networks. Neural Netw 163:298--311. \doi{10.1016/j.neunet.2023.04.006}

\bibitem[{Kohonen(1982)}]{kohonen1982}
Kohonen T (1982) Self-organized formation of topologically correct feature maps. Biol Cybern 43(1):59--69. \doi{10.1007/BF00337288}

\bibitem[{Latora and Marchiori(2001)}]{latora2001}
Latora V, Marchiori M (2001) Efficient behavior of small-world networks. Phys Rev Lett 87:198701. \doi{10.1103/PhysRevLett.87.198701}

\bibitem[{Laughlin and Sejnowski(2003)}]{laughlin2003communication}
Laughlin SB, Sejnowski TJ (2003) Communication in neuronal networks. Science 301(5641):1870--1874. \doi{10.1126/science.1089662}

\bibitem[{Linsker(1988)}]{linsker1988}
Linsker R (1988) Self-organization in a perceptual network. Computer 21(3):105--117. \doi{10.1109/2.36}

\bibitem[{Lord et~al.(2017)Lord, Stevner, Deco, and Kringelbach}]{lord2017understanding}
Lord LD, Stevner AB, Deco G, et~al (2017) Understanding principles of integration and segregation using whole-brain computational connectomics: implications for neuropsychiatric disorders. Phil Trans R Soc A 375(2096):20160283. \doi{10.1098/rsta.2016.0283}

\bibitem[{Lorenz(1963)}]{lorenz1963}
Lorenz EN (1963) Deterministic nonperiodic flow. J Atmos Sci 20(2):130--141. \doi{10.1175/1520-0469(1963)020<0130:DNF>2.0.CO;2}

\bibitem[{Von~der Malsburg(1973)}]{malsburg1973}
Von~der Malsburg C (1973) Self-organization of orientation sensitive cells in the striate cortex. Kybernetik 14(2):85--100. \doi{10.1007/BF00288907}

\bibitem[{Newman(2018)}]{newman2018networks}
Newman M (2018) Networks. OUP Oxford

\bibitem[{Newman(2004)}]{newman2004analysis}
Newman ME (2004) Analysis of weighted networks. Phys Rev E 70(5):056131. \doi{10.1103/PhysRevE.70.056131}

\bibitem[{Rigotti et~al.(2013)Rigotti, Barak, Warden, Wang, Daw, Miller, and Fusi}]{rigotti2013importance}
Rigotti M, Barak O, Warden MR, et~al (2013) The importance of mixed selectivity in complex cognitive tasks. Nature 497:585--590. \doi{10.1038/nature12160}

\bibitem[{Rolls(2013)}]{rolls2013}
Rolls ET (2013) The mechanisms for pattern completion and pattern separation in the hippocampus. Front Syst Neurosci 7:74. \doi{10.3389/fnsys.2013.00074}

\bibitem[{R\"ossler(1976)}]{rossler1976}
R\"ossler O (1976) An equation for continuous chaos. Phys Lett A 57(5):397--398. \doi{https://doi.org/10.1016/0375-9601(76)90101-8}

\bibitem[{Serra et~al.(2018)Serra, Suris, Miron, and Karatzoglou}]{serra2018overcoming}
Serra J, Suris D, Miron M, et~al (2018) Overcoming catastrophic forgetting with hard attention to the task. In: Dy J, Krause A (eds) Proceedings of the 35th International Conference on Machine Learning, Proceedings of Machine Learning Research, vol~80. PMLR, pp 4548--4557, \urlprefix\url{https://proceedings.mlr.press/v80/serra18a.html}

\bibitem[{Song et~al.(2016)Song, Yang, and Wang}]{song2016training}
Song HF, Yang GR, Wang XJ (2016) Training excitatory-inhibitory recurrent neural networks for cognitive tasks: a simple and flexible framework.  PLoS Comput Biol 12(2):e1004792. \doi{10.1371/journal.pcbi.1004792}

\bibitem[{Sporns(2016)}]{sporns2016book}
Sporns O (2016) Networks of the Brain. MIT press

\bibitem[{Sporns and Betzel(2016)}]{sporns2016}
Sporns O, Betzel RF (2016) Modular brain networks. Annu Rev Psychol 67:613--640. \doi{10.1146/annurev-psych-122414-033634}

\bibitem[{Sussillo(2014)}]{sussillo2014neural}
Sussillo D (2014) Neural circuits as computational dynamical systems. Curr Opin Neurobiol 25:156--163. \doi{10.1016/j.conb.2014.01.008}

\bibitem[{Sussillo and Abbott(2009)}]{sussillo2009}
Sussillo D, Abbott LF (2009) Generating coherent patterns of activity from chaotic neural networks. Neuron 63(4):544--557. \doi{10.1016/j.neuron.2009.07.018}

\bibitem[{Tanaka et~al.(2009)Tanaka, Kaneko, and Aoyagi}]{tanaka2009}
Tanaka T, Kaneko T, Aoyagi T (2009) {Recurrent infomax generates cell assemblies, neuronal avalanches, and simple cell-like selectivity.} Neural Comput 21(4):1038--67. \doi{10.1162/neco.2008.03-08-727}

\bibitem[{Tsuda et~al.(2016)Tsuda, Yamaguti, and Watanabe}]{tsuda2016}
Tsuda I, Yamaguti Y, Watanabe H (2016) Self-organization with constraints-a mathematical model for functional differentiation. Entropy 18(3):74. \doi{10.3390/e18030074}

\bibitem[{Tsuda et~al.(2022)Tsuda, Watanabe, Tsukada, and Yamaguti}]{tsuda2022nature}
Tsuda I, Watanabe H, Tsukada H, et~al (2022) On the nature of functional differentiation: The role of self-organization with constraints. Entropy 24(2):240. \doi{10.3390/e24020240}

\bibitem[{Tsukada et~al.(1996)Tsukada, Aihara, Saito, and Kato}]{tsukada1996}
Tsukada M, Aihara T, Saito H, Kato H (1996) Hippocampal LTP depends on spatial and temporal correlation of inputs. Neural Networks 9:1357--1365. \doi{10.1016/s0893-6080(96)00047-0}

\bibitem[{Tsukada and Pan(2005)}]{tsukada2005}
Tsukada M, Pan X (2005) The spatiotemporal learning rule and its efficiency in separating spatiotemporal patterns. Biological Cybernetics 92:139--146. \doi{10.1007/s00422-004-0523-1}

\bibitem[{Veniat et~al.(2021)Veniat, Denoyer, and Ranzato}]{veniat2021efficient}
Veniat T, Denoyer L, Ranzato M (2021) Efficient continual learning with modular networks and task-driven priors. In: 9th International Conference on Learning Representations, ICLR 2021

\bibitem[{Watanabe et~al.(2020)Watanabe, Ito, and Tsuda}]{watanabe2020mathematical}
Watanabe H, Ito T, Tsuda I (2020) A mathematical model for neuronal differentiation in terms of an evolved dynamical system. Neurosci Res 156:206--216. \doi{10.1016/j.neures.2020.02.003}

\bibitem[{Yamaguti and Tsuda(2015)}]{yamaguti2015mathematical}
Yamaguti Y, Tsuda I (2015) Mathematical modeling for evolution of heterogeneous modules in the brain. Neural Netw 62:3--10. \doi{10.1016/j.neunet.2014.07.013}

\bibitem[{Yamaguti and Tsuda(2021)}]{yamaguti2021functional}
Yamaguti Y, Tsuda I (2021) Functional differentiations in evolutionary reservoir computing networks. Chaos 31(1):013137. \doi{10.1063/5.0019116}

\bibitem[{Yang et~al.(2019)Yang, Joglekar, Song, Newsome, and Wang}]{yang2019task}
Yang GR, Joglekar MR, Song HF, et~al (2019) Task representations in neural networks trained to perform many cognitive tasks. Nat Neurosci 22(2):297--306. \doi{10.1038/s41593-018-0310-2}

\end{thebibliography}

\end{document}